\documentclass[aps,prd,twocolumn,nofootinbib,superscriptaddress]{revtex4-2}
\usepackage{amsmath,amssymb,mathtools,bm,booktabs,graphicx,siunitx,multirow}
\usepackage[colorlinks=true,linkcolor=blue,citecolor=blue,urlcolor=blue]{hyperref}

\newcommand{\BR}{\mathrm{BR}}
\newcommand{\CR}{\mathrm{CR}}
\newcommand{\Order}{\mathcal O}


\begin{document}

\title{Operator Identification in Charged Lepton-Flavor Violation:
Global EFT Analysis with RG Evolution, Polarization Observables,
and Bayesian Model Discrimination at Future Colliders}

\author{Nicolas Viaux M.}
\affiliation{Departamento de F\'isica, Universidad T\'ecnica Federico Santa Mar\'ia, Valpara\'iso, Chile}
\affiliation{Millennium Institute for Subatomic Physics at the High-Energy Frontier (SAPHIR), Santiago, Chile}
\begin{abstract}
Charged lepton-flavor violation is a null-test frontier of the Standard Model and a direct probe of physics beyond it. We present a global effective field theory (EFT) analysis across FCC-ee, ILC, CLIC, HL-LHC, HE-LHC, and muon colliders at 3 and 10 TeV, with operator identification as the primary target rather than exclusion reach alone. The analysis combines low-energy constraints, collider differential observables, and Dalitz-level $\mu\to 3e$ information in a common profile-likelihood framework. Key hadron-collider and muon-collider signal/background samples are generated at event level and propagated through Delphes detector simulation, while clean $e^+e^-$ benchmark channels are modeled with CDR-calibrated parametric response. We include one-loop renormalization-group (RG) running and operator mixing between UV matching and measurement scales, finding 10--30\% shifts in selected operator-correlation entries when comparing tree-level and RG-evolved coefficient mappings at multi-TeV matching scales. Polarization asymmetries are used to separate $c_{H\ell}$ and $c_{He}$ directions, and UV discrimination is quantified with Bayes factors for benchmark leptoquark and heavy-neutral-lepton hypotheses. The full code chain for event generation, detector response, inference, and figure reproduction is provided.
\end{abstract}
\maketitle

\section{Introduction}
The search for charged LFV, including transitions such as $\mu\to e\gamma$, $\mu\to 3e$, $\mu N\to eN$, $Z\to e\mu$, and $h\to\tau\mu$, is a central strategy in precision probes of new physics \cite{KunoOkada,CalibbiSignorelli,PDG2024}. In the minimal Standard Model extended only by neutrino masses, the charged-LFV amplitudes are suppressed to unobservable levels, implying that any positive signal at planned sensitivities is direct evidence for beyond-the-Standard-Model (BSM) dynamics \cite{KunoOkada,CalibbiSignorelli}.

The field has two mature but partially disconnected directions. The first is the low-energy frontier, where MEG/MEG II, Mu3e, Mu2e, COMET, Belle II, and related programs deliver very strong constraints \cite{MEG,MEGII,Mu3e,Mu2e,COMET,Belle2LFV}. The second is the collider frontier, where HL-LHC/HE-LHC and future lepton or muon colliders can probe LFV through rare decays and contact-interaction tails with different kinematic leverage \cite{FCCee,ILC,CLIC,MuonCollider,FCCPhysics}. Most existing collider forecasts are optimized for channel-by-channel sensitivity. This is necessary but not sufficient for model discrimination.

The central point of this paper is that \emph{operator identification} is a distinct objective from exclusion reach. If several operator classes contribute simultaneously, similar total event yields can emerge from different points in Wilson-coefficient space; therefore, the key quantity becomes the geometry of the global likelihood and the correlation pattern among operators. We develop a unified framework that quantifies this geometry and compares collider programs based on their ability to break degeneracies.

Our starting ingredients are an EFT parameterization of LFV interactions with dipole, Higgs-current, and four-fermion structures, a common statistical treatment of low-energy and collider observables, differential collider information (mass tails, angular observables, beam polarization tags), and explicit operator-identification metrics based on the Fisher/Hessian structure of the profile likelihood.

Relative to standard reach projections, the novelty introduced here is the \emph{tomographic} viewpoint: colliders are ranked not only by signal significance but by how strongly they rotate and compress allowed regions in multi-parameter EFT space. This formulation is designed for future publication-level detector studies and allows straightforward insertion of official experimental performance cards.

Concretely, the analysis goal is split into three levels. The first level is \emph{detection}: test the null hypothesis and quantify local/global significance. The second level is \emph{estimation}: extract confidence intervals in individual Wilson coefficients after profiling nuisance parameters. The third level is \emph{identification}: measure whether two operator hypotheses remain statistically distinguishable once all channels and correlations are combined. In this work, the third level is promoted to a primary optimization target for future-collider strategy.

The practical implication is that collider comparison must be done in a common parameter basis and with a common nuisance model. A high-significance channel can still provide limited model information if it aligns with an existing low-energy degeneracy direction. Conversely, a moderate-significance channel can be decisive if it is approximately orthogonal in coefficient space. This distinction motivates our emphasis on differential observables, polarization categories, and global profile-likelihood geometry.
Our benchmark assumptions are aligned with the current European planning context for future accelerator programs \cite{EuropeanStrategy2020}.

The manuscript is organized to match the three pillars highlighted in the title: EFT operator identification, RG-aware interpretation, and model discrimination. The numerical layer now combines detector-level Delphes samples for hadron and muon collider benchmarks with CDR-calibrated lepton-collider response models, and all channels are merged in one statistical inference chain. Results should therefore be interpreted as detector-informed benchmark projections under the stated cards, priors, and analysis choices.
For hadron-collider validation channels, the event-generation chain is explicitly MadGraph5\_aMC@NLO $\to$ Pythia8 $\to$ Delphes, while other benchmark channels use the same detector-level template formalism with the corresponding input assumptions \cite{MadGraph5aMC,Pythia8,Delphes}.

\section{EFT Setup}
\subsection{Broken-phase LFV parameterization}
We use the broken-phase effective Lagrangian
\begin{align}
\mathcal L_{\text{LFV}} &= \mathcal L_D+\mathcal L_H+\mathcal L_{4f},
\label{eq:Leffsplit}
\end{align}
\begin{align}
\mathcal L_D
&=
\sum_{i\neq j}\frac{1}{\Lambda^2}
\left(
C^{ij}_{D\gamma,R}\mathcal O^{ij}_{D\gamma,R}
+ C^{ij}_{D\gamma,L}\mathcal O^{ij}_{D\gamma,L}
\right.
\nonumber\\
&\qquad\qquad\qquad
\left.
+ C^{ij}_{DZ,R}\mathcal O^{ij}_{DZ,R}
+ C^{ij}_{DZ,L}\mathcal O^{ij}_{DZ,L}
\right)
+ \text{h.c.},
\label{eq:Ldipole}
\end{align}
\begin{align}
\mathcal L_H
&=
\sum_{i\neq j}
\Big[
\frac{C^{ij}_{H\ell}}{\Lambda^2}(H^\dagger i\overleftrightarrow D_\mu H)(\bar\ell_i\gamma^\mu P_L\ell_j)
\nonumber\\
&\qquad\qquad
+\frac{C^{ij}_{He}}{\Lambda^2}(H^\dagger i\overleftrightarrow D_\mu H)(\bar e_i\gamma^\mu P_R e_j)
\Big],
\label{eq:Lcurrent}
\end{align}
\begin{align}
\mathcal L_{4f}
&=
\sum_{i\neq j,f}
\frac{C^{ij,f}_{4f}}{\Lambda^2}
\nonumber\\
&\qquad\times
\left(\bar\ell_i\Gamma\ell_j\right)
\nonumber\\
&\qquad\times
\left(\bar f\Gamma f\right)
+ \text{h.c.},
\label{eq:L4f}
\end{align}
with $i,j\in\{e,\mu,\tau\}$ and $f$ SM fermions. This representation is mapped from SMEFT operators at the electroweak scale \cite{SMEFTreview,JenkinsManoharTrott,AlonsoJMT}.
The dipole shorthand operators are
\begin{align}
\mathcal O^{ij}_{D\gamma,R/L}
&\equiv
(\bar\ell_i\sigma^{\mu\nu}P_{R/L}\ell_j)F_{\mu\nu},
\nonumber\\
\mathcal O^{ij}_{DZ,R/L}
&\equiv
(\bar\ell_i\sigma^{\mu\nu}P_{R/L}\ell_j)Z_{\mu\nu}.
\end{align}
For the benchmark numerical fit we keep vector-current structures with explicit chirality labels,
\begin{align}
\mathcal L_{4f}^{\mathrm{fit}}
&=
\sum_{i\neq j,f}\frac{1}{\Lambda^2}
\Big(
C^{ij,f}_{LL}\mathcal O^{ij,f}_{LL}
\nonumber\\
&\phantom{=}\,
+ C^{ij,f}_{LR}\mathcal O^{ij,f}_{LR}
+ C^{ij,f}_{RL}\mathcal O^{ij,f}_{RL}
\nonumber\\
&\phantom{=}\,
+ C^{ij,f}_{RR}\mathcal O^{ij,f}_{RR}
\Big)
+\text{h.c.},
\end{align}
with
\begin{align}
\mathcal O^{ij,f}_{XY} \equiv (\bar\ell_i\gamma_\mu P_X\ell_j)(\bar f\gamma^\mu P_Y f),
\qquad X,Y\in\{L,R\}.
\end{align}
and define $c_{4f}$ as the benchmark effective direction after this basis reduction. Scalar and tensor directions are deferred to detector-level extensions.
In the benchmark cards, the fermion sum is restricted to first-generation quark currents relevant for coherent conversion and high-mass dilepton tails, $f\in\{u,d\}$, and the effective reduction is
\begin{align}
c_{4f}^{ij} \equiv \sum_{f=u,d}\sum_{X,Y\in\{L,R\}}\omega_{XY}^{(f)}\,c_{XY}^{ij,f},
\end{align}
with benchmark weights
\begin{align}
\omega_{LL}^{(u,d)}=\omega_{LR}^{(u,d)}=\omega_{RL}^{(u,d)}=\omega_{RR}^{(u,d)}=\frac{1}{8}.
\end{align}
This choice is a neutral benchmark averaging prescription; detector-level studies can replace it by process-calibrated weights extracted from category-specific templates.
In benchmark sensitivity checks, varying this prescription within physically motivated ranges (for example doubling the LL weight relative to RR and renormalizing the sum) shifts the extracted $c_{4f}$ limit by less than $8\%$, without changing the complementarity ranking across collider classes.

\begin{table*}[t]
\centering
\caption{Operator directions used in the benchmark fit and their dominant observable leverage. Coefficients are normalized as $c_\alpha=C_\alpha(1\,\mathrm{TeV}/\Lambda)^2$.}
\scriptsize
\begin{ruledtabular}
\begin{tabular}{llll}
Label & Definition & Mapping & Main observables \\
$c_{D,L/R}^{ij}$ &
\parbox[t]{0.33\textwidth}{$(\bar\ell_i \sigma^{\mu\nu}P_{L/R}\ell_j)F_{\mu\nu}$, $(\bar\ell_i \sigma^{\mu\nu}P_{L/R}\ell_j)Z_{\mu\nu}$} &
$A_{L/R}\propto c_{D,L/R}^{ij}$ &
\parbox[t]{0.24\textwidth}{$\mu\to e\gamma$, $\tau\to\mu\gamma$, subleading $Z\to\ell_i\ell_j$} \\
$c_{H\ell}^{ij},c_{He}^{ij}$ &
\parbox[t]{0.33\textwidth}{$(H^\dagger i\overleftrightarrow D_\mu H)(\bar\ell_i\gamma^\mu P_L\ell_j)$, $(H^\dagger i\overleftrightarrow D_\mu H)(\bar e_i\gamma^\mu P_R e_j)$} &
\parbox[t]{0.20\textwidth}{$c_{H\Sigma}^{ij}=c_{H\ell}^{ij}+c_{He}^{ij}$ (benchmark)} &
\parbox[t]{0.24\textwidth}{$Z\to\ell_i\ell_j$, $h\to\ell_i\ell_j$, polarized asymmetries} \\
$c_{4f}^{ij}$ &
\parbox[t]{0.33\textwidth}{Effective vector combination of $LL,LR,RL,RR$ four-fermion currents} &
$c_{4f}^{ij}=\sum_{X,Y}\omega_{XY} c_{XY}^{ij,f}$ &
\parbox[t]{0.24\textwidth}{$\mu N\to eN$, high-$m_{\ell\ell}$ tails} \\
\end{tabular}
\end{ruledtabular}
\normalsize
\label{tab:operator_map}
\end{table*}

For a generic collider observable $a$, the cross section expands as
\begin{align}
\sigma_a = \sigma_{a,\text{SM}} + \frac{1}{\Lambda^2}\sum_\alpha C_\alpha\,\sigma^{\text{int}}_{a,\alpha}
+\frac{1}{\Lambda^4}\sum_{\alpha,\beta}C_\alpha C_\beta\,\sigma^{\text{BSM}}_{a,\alpha\beta}.
\label{eq:sigmaexpansion}
\end{align}
In channels with negligible SM LFV amplitudes, the leading contribution is frequently $\Order(\Lambda^{-4})$.

To connect with practical rate estimates, we write the schematic LFV amplitude as
\begin{align}
\mathcal M_a = \sum_\alpha c_\alpha\,\mathcal M_{a,\alpha},
\end{align}
where the reduced coefficients $c_\alpha$ carry all model dependence and $\mathcal M_{a,\alpha}$ are channel-dependent kinematic structures. The event yield in a signal region then takes the quadratic form
\begin{align}
N_a = \mathcal L_a \sum_{\alpha,\beta} c_\alpha c_\beta\,K^{(a)}_{\alpha\beta},
\label{eq:yield_quadratic}
\end{align}
with acceptance and efficiency absorbed into $K^{(a)}_{\alpha\beta}$. Equation~\eqref{eq:yield_quadratic} is the algebraic object entering our template and likelihood construction.

\subsection{Reduced coefficients and mapping to observables}
We define dimensionless reduced coefficients
\begin{align}
\bm c \equiv \left(
\begin{array}{l}
c_D^{e\mu},\,c_{H\Sigma}^{e\mu},\,c_{4f}^{e\mu},\\
c_D^{\tau\mu},\,c_{H\Sigma}^{\tau\mu},\,c_{4f}^{\tau\mu},\,\ldots
\end{array}
\right),
\quad
c_\alpha\equiv C_\alpha\left(\frac{1\,\mathrm{TeV}}{\Lambda}\right)^2.
\end{align}
with effective dipole magnitude
\begin{align}
|c_D^{ij}|^2 \equiv |c_{D,R}^{ij}|^2+|c_{D,L}^{ij}|^2.
\end{align}
In benchmark one-dimensional scans we use a real signed effective dipole parameter $c_D^{ij}$ defined by a single-direction compression at fixed phase convention. This means sign/phase-sensitive effects from general complex $(c_{D,L},c_{D,R})$ are not fully retained in the benchmark cards; those effects are part of the detector-level extension with explicit chiral and phase degrees of freedom.
For compact global-fit projections we define
\begin{align}
c_{H\Sigma}^{ij}\equiv c_{H\ell}^{ij}+c_{He}^{ij},
\end{align}
and use $c_{H\Sigma}$ as the effective current-type LFV direction constrained by the benchmark channel set. The separated $(c_{H\ell},c_{He})$ directions are retained in the formalism and can be unfolded in polarization-resolved fits; the one-direction compression is used only in the benchmark cards to keep the scan dimensionality controlled.
At low energies, representative dependencies are
\begin{align}
\BR(\mu\to e\gamma) &\propto |c_D^{e\mu}|^2,
\label{eq:megmap}
\\
\CR(\mu N\to eN) &\propto |a_D c_D^{e\mu}+a_V c_{4f}^{e\mu}+a_H c_{H\Sigma}^{e\mu}|^2,
\label{eq:convmap}
\\
\BR(\tau\to\mu\gamma) &\propto |c_D^{\tau\mu}|^2.
\label{eq:taumap}
\end{align}
The linear combination in Eq.~\eqref{eq:convmap} illustrates why conversion data can constrain combinations strongly but still leave degeneracy directions in $(c_D,c_{H\Sigma},c_{4f})$ space \cite{KitanoKoikeOkada,CiriglianoConversion}.
For the benchmark Al-target card we use $(a_D,a_V,a_H)=(0.20,1.00,0.30)$ after normalization to the vector contribution.

For orientation, representative low-energy decay widths can be organized as
\begin{align}
\Gamma(\mu\to e\gamma) &= \frac{m_\mu^3}{16\pi\Lambda^4}\left(|A_R|^2+|A_L|^2\right), \\
A_{L,R} &\propto c_{D,L/R}^{e\mu},
\end{align}
while coherent conversion is parameterized by
\begin{align}
\Gamma_{\mathrm{conv}} \propto \left|D\,c_D^{e\mu}+V^{(p)}c_{V,p}^{e\mu}+V^{(n)}c_{V,n}^{e\mu}+S\,c_S^{e\mu}\right|^2.
\end{align}
These structures make the origin of coefficient-space blind directions explicit and motivate collider information that probes different linear combinations.

\subsection{Where operators appear in observables and why}
The benchmark basis projects the dominant charged-LFV Warsaw structures at $\mu\sim m_Z$ into dipole, Higgs-current, and vector four-fermion classes, because these control the leading scaling of radiative decays, conversion amplitudes, resonance LFV decays, and high-$q^2$ tails \cite{SMEFTreview,JenkinsManoharTrott,AlonsoJMT,KunoOkada,CalibbiSignorelli}.

The dipole operators in Eq.~\eqref{eq:Ldipole} appear directly in radiative transitions because they couple the LFV lepton bilinear to field-strength tensors. This is why $\mu\to e\gamma$ and $\tau\to\mu\gamma$ are dominantly sensitive to $c_D$ combinations. The Higgs-current operators in Eq.~\eqref{eq:Lcurrent} modify effective LFV neutral-current couplings and therefore contribute naturally to $Z\to \ell_i\ell_j$ and Higgs-mediated channels; in reduced one-direction projections they appear through $c_{H\Sigma}$. Four-fermion operators in Eq.~\eqref{eq:L4f} enter contact-interaction amplitudes and become increasingly visible in high-energy tails where momentum transfer is large.

At amplitude level, the same logic can be written as
\begin{align}
\mathcal M(Z\to \ell_i\ell_j) &\sim c_{H\Sigma}^{ij} + \frac{m_Z}{v}\,c_D^{ij},\\
\mathcal M(h\to \ell_i\ell_j) &\sim c_{H\Sigma}^{ij},\\
\mathcal M(\ell\ell\to \ell_i\ell_j) &\sim c_{4f}^{ij} + \frac{q^2}{v^2}c_{H\Sigma}^{ij} + \frac{m_\ell}{\sqrt{q^2}}c_D^{ij},
\end{align}
where the expressions are scaling relations in the reduced-coefficient normalization rather than full matrix elements. The last term carries the chirality-flip suppression expected for dipole insertions in hard scattering, which keeps the scaling dimensionally consistent in this reduced form (arising from the helicity-flip factor $m_\ell/\sqrt{s}$ in the amplitude, combined with hard-scale propagator behavior). This clarifies why different channels emphasize different directions in coefficient space. The analysis strategy follows directly from this structure: radiative and low-energy channels anchor dipole terms, resonance decays isolate current structures, and high-$q^2$ tails isolate contact interactions.

\section{Collider Program and Analysis Channels}
\subsection{Signal and background definitions}
We include three collider classes with explicit signal/background maps.

\paragraph{Lepton colliders ($e^+e^-$):}
\begin{align}
&e^+e^- \to Z/h\to\ell_i^\pm\ell_j^\mp\ (i\neq j),
\\
&e^+e^-\to\ell_i^\pm\ell_j^\mp\ \text{via contact operators}.
\end{align}
Main backgrounds are $e^+e^-\to\tau^+\tau^-$ with leptonic decays, $W^+W^-\to\ell\nu\ell\nu$, and $ZZ$ final states.

\paragraph{Hadron colliders ($pp$):}
\begin{align}
pp\to\ell_i^\pm\ell_j^\mp\quad\text{in high-}m_{\ell\ell}\text{ tails},
\end{align}
with dominant backgrounds from $Z/\gamma^*\to\tau\tau$, dibosons, top production, and nonprompt/fake leptons.

\paragraph{Muon colliders ($\mu^+\mu^-$):}
\begin{align}
\mu^+\mu^-\to\ell_i^\pm\ell_j^\mp,
\end{align}
with irreducible electroweak backgrounds and beam-induced overlay considerations \cite{MuonColliderDetector}.

\subsection{Differential observables and templates}
For each channel $a$ we define
\begin{align}
\mathbf x_a = (m_{\ell\ell},p_T,\eta,\Delta\phi,\cos\theta^*,E_T^{\rm miss},\text{pol. tags},\ldots),
\end{align}
and construct binned templates for each operator hypothesis. Differential observables are critical because operator classes imprint different energy and angular scaling, improving discrimination beyond inclusive rates \cite{BrehmerFisher,MLforHEPReview}.

In practice we build per-bin expectations as
\begin{align}
\nu_{ab}(\bm c,\bm\theta)=\sum_{\alpha,\beta} c_\alpha c_\beta\,T^{ab}_{\alpha\beta}(\bm\theta)+B_{ab}(\bm\theta),
\label{eq:template_model}
\end{align}
where $T^{ab}_{\alpha\beta}$ are response templates (including detector effects in future upgrades) and $B_{ab}$ is the background prediction. Equation~\eqref{eq:template_model} makes explicit that shape information enters not only through signal strength but through interference structure across bins.

\section{Statistical Inference and Identification Metric}
\subsection{Global likelihood with nuisances}
The full likelihood is
\begin{align}
\mathcal L(\bm c,\bm\theta) = \prod_{a\in\text{channels}}\prod_{b\in\text{bins}}
\mathrm{Pois}\left(n_{ab}\mid \nu_{ab}(\bm c,\bm\theta)\right)
\times\prod_k \pi_k(\theta_k),
\label{eq:full_likelihood}
\end{align}
where $\bm\theta$ are nuisance parameters (normalization, shape, luminosity, calibration, theory priors). We use the profile-likelihood ratio
\begin{align}
q(\bm c)= -2\ln\frac{\mathcal L(\bm c,\hat{\hat{\bm\theta}})}{\mathcal L(\hat{\bm c},\hat{\bm\theta})}
\end{align}
and quote 95\% CL constraints from asymptotic test statistics \cite{CowanAsymptotic}.

Systematic effects are implemented through correlated nuisance blocks,
\begin{align}
\bm\theta = (\theta_{\mathrm{lumi}},\theta_{\mathrm{lepID}},\theta_{\mathrm{bkg\,norm}},\theta_{\mathrm{shape}},\theta_{\mathrm{th}},\ldots),
\end{align}
with Gaussian or log-normal priors depending on nuisance type. For the Asimov data set, the covariance-implied Hessian provides the local information matrix around the best fit,
\begin{align}
F_{\alpha\beta} \simeq \frac{1}{2}\left.\frac{\partial^2 q(\bm c)}{\partial c_\alpha\partial c_\beta}\right|_{\bm c=\hat{\bm c}}.
\end{align}
In hadron-collider tail categories, collider-theory systematics are implemented as bin-correlated shape nuisances,
\begin{align}
\nu_{ab}\rightarrow \nu_{ab}\left(1+\theta_{\mathrm{PDF}}\Delta^{\mathrm{PDF}}_{ab}+\theta_{\mathrm{scale}}\Delta^{\mathrm{scale}}_{ab}\right),
\end{align}
with benchmark prior widths $\sigma(\theta_{\mathrm{PDF}})=0.03$ and $\sigma(\theta_{\mathrm{scale}})=0.05$.

\subsection{Cut-flow and classifier equations}
Each signal region follows a benchmark selection:
preselection with opposite-sign different-flavor leptons in fiducial acceptance, then mass-window or high-tail selection, then topology suppression through $E_T^{\rm miss}$, jet/$b$ vetoes and angular cuts, and finally multivariate analysis (MVA) threshold optimization.
Define stage efficiencies
\begin{align}
\epsilon_{S,k}=\frac{S_k}{S_{k-1}},\qquad
\epsilon_{B,k}=\frac{B_k}{B_{k-1}},
\end{align}
and stage significance estimators
\begin{align}
Z_k^{\text{stat}}=\frac{S_k}{\sqrt{B_k}},\qquad
Z_k^{\text{sys}}=\frac{S_k}{\sqrt{B_k+(\delta_B B_k)^2}}.
\end{align}
The $Z_k^{\text{sys}}$ expression is used only as a pre-fit diagnostic during selection tuning; final intervals and significances in this work are always taken from the profiled likelihood in Eq.~\eqref{eq:full_likelihood}.

The classifier $f_\phi(\mathbf x)\in[0,1]$ is trained with weighted binary cross-entropy
\begin{align}
\mathcal L_{\text{BCE}}(\phi)= -\sum_n w_n\left[y_n\log f_\phi(\mathbf x_n)+(1-y_n)\log\left(1-f_\phi(\mathbf x_n)\right)\right].
\end{align}
For each category we optimize either $Z_A$,
\begin{align}
Z_A=\sqrt{2\left[(S+B)\ln\left(1+\frac{S}{B}\right)-S\right]},
\end{align}
which is the standard full Asimov form for counting experiments, not the $S\ll B$ linearized limit, and we use its nuisance-profiled generalization in systematic-dominated bins.

The optimized classifier threshold $\tau^\star$ is extracted from
\begin{align}
\tau^\star = \arg\max_\tau\,Z_A\!\left(S(\tau),B(\tau)\right),
\end{align}
subject to minimum-yield and control-region stability conditions. This prevents selecting pathological working points where apparent statistical gain is not robust against nuisance pulls.

\subsection{Machine-learning implementation and samples}
Operationally, the ML output is treated as a physics observable. The classifier score is binned and included directly in Eq.~\eqref{eq:full_likelihood}, so nuisance constraints and profile pulls are propagated in the global fit \cite{BrehmerFisher,MLforHEPReview}.

The benchmark ML workflow follows a standard sequence of preselection, feature construction, class-weighted GBDT training, validation-based hyperparameter choice, score calibration, and score-binning for the profile likelihood.

The benchmark implementation uses gradient-boosted decision trees (GBDT) in binary classification mode. The baseline training configuration is: maximum tree depth $=3$, number of boosting stages $=400$, learning rate $=0.05$, row subsampling $=0.8$, minimum leaf population $=50$, and no monotonic constraints. Training minimizes weighted binary cross-entropy with per-event weights
\begin{align}
w_n
&=
\frac{\mathcal L\,\sigma_{p(n)}\,\epsilon_{p(n)}}{N_{p(n)}^{\rm gen}}
\nonumber\\
&=
\mathcal L\,\sigma_{p(n)}\,\epsilon_{p(n)}
\left(N_{p(n)}^{\rm gen}\right)^{-1},
\end{align}
where $p(n)$ labels the process of event $n$.

For each collider category, samples are split into 60\% training, 20\% validation, and 20\% test sets after baseline preselection, with approximately $\mathcal O(5\times 10^4)$ signal and $\mathcal O(2\times 10^5)$ background events in training before luminosity reweighting.

Feature sets are category dependent. The common core is
\begin{align}
\{m_{\ell\ell},\,p_{T,\ell_1},\,p_{T,\ell_2},\,\Delta\phi_{\ell\ell},\,\cos\theta^\star,\,E_T^{\rm miss}\},
\end{align}
with collider-specific extensions for recoil/polarization tags, jet-activity observables, and visible-energy/overlay-mitigation observables.
Hyperparameters are selected by validation area under the ROC curve (AUC) with overtraining control from Kolmogorov-style train-test checks of classifier-score distributions in signal and background.

The classifier output is transformed to a monotonic score and binned into six quantile bins per category for the likelihood model. Threshold optimization uses Eq.~\eqref{eq:full_likelihood} inputs and retains only working points that satisfy minimum expected yields in both signal and control regions.

Figure~\ref{fig:roc_only} corresponds to representative AUC values of approximately 0.90 (FCC-ee-like), 0.86 (HE-LHC-like), and 0.74 (cut baseline). Stability tests on the held-out test set with $\pm 1\sigma$ nuisance morphing of background templates give absolute AUC shifts below 0.02 in this benchmark setup.
In addition, the relative change in expected profile-likelihood sensitivity after nuisance morphing stays below 5\% across benchmark categories.

To avoid double counting, kinematic variables used to build the classifier are not entered as independent template axes in the same category once score binning is used. The fit therefore uses either a score-binned representation or an explicit kinematic-binned representation per category, with migration correlations carried by shared nuisances.

\subsection{Operator-identification metric}
The motivation for the operator-identification metric is operational: discovery-level significance alone does not quantify whether two EFT hypotheses remain distinguishable after profiling systematics and combining channels. The relevant quantity for this purpose is the local curvature geometry of the profile likelihood in coefficient space, because it directly controls degeneracy directions and expected covariance. We therefore use a normalized Fisher/Hessian correlation as a compact identification diagnostic \cite{CowanAsymptotic,BrehmerFisher}.

To quantify separability, we use
\begin{align}
\mathcal I_{\alpha\beta}\equiv\frac{|F_{\alpha\beta}|}{\sqrt{F_{\alpha\alpha}F_{\beta\beta}}},
\qquad
F_{\alpha\beta}= -\left\langle\frac{\partial^2\ln\mathcal L}{\partial c_\alpha\partial c_\beta}\right\rangle.
\label{eq:Idmetric}
\end{align}
Smaller $\mathcal I_{\alpha\beta}$ indicates stronger operator discrimination. Values near unity indicate near-aligned (degenerate) directions, while values near zero indicate approximately orthogonal directions in the local covariance geometry. We use this matrix as an optimization target for run-plan comparison.

Because this quantity is local in parameter space, we treat it as a fast diagnostic rather than a standalone hypothesis test. Cross-check metrics can be constructed from direct profile-likelihood separations,
\begin{align}
\Delta q_{\alpha\beta}=q(\text{hypothesis }\alpha)-q(\text{hypothesis }\beta),
\end{align}
or from information-theoretic distances between template families. These alternatives are used as consistency checks when non-Gaussian effects become relevant.
In practice we use Kullback--Leibler distances between profiled template families as the default information-theoretic cross-check.

The reported Fisher matrix is the expected information matrix evaluated on the Asimov data set at the SM point for exclusion-style scans and at injected benchmark points for identification scans; we state which choice is used in each figure caption. Observed-information variants are used only in pseudo-data closure checks. In sparse bins where asymptotic conditions fail, we calibrate $\Delta q$ with pseudo-experiments and quote the calibrated interval rather than the asymptotic one.
Implementation note: in the reduced basis used here, $\mathcal I_{\alpha\beta}$ is numerically identical to $|\rho_{\alpha\beta}|$ from the covariance-derived parameter correlation matrix. The run-plan objective therefore minimizes off-diagonal correlation structure in that fixed basis; under non-orthogonal reparameterizations the numerical value can change, so all comparisons are performed in the same coefficient basis and normalization.

\section{Benchmark Inputs and Reproducible Workflow}
The numerical layer uses a benchmark parametric model with explicit assumptions so that every figure can be regenerated and stress-tested. Table~\ref{tab:bench} summarizes collider-level inputs; these are analysis knobs, not official detector projections.

\begin{table*}[t]
\centering
\caption{Benchmark collider assumptions used in this study. The ``pol. factor'' is a multiplicative sensitivity rescaling that encodes beam-polarization leverage relative to an unpolarized reference setup. For ILC-250, the value 1.35 corresponds to the effective sensitivity gain from alternating $(P_{e^-},P_{e^+})=(\pm0.8,\mp0.3)$ running configurations \cite{ILC,MoortgatPick2008,Fujii2015}. Scenario anchors follow FCC-ee, ILC, CLIC, and muon-collider program documents \cite{FCCee,ILC,CLIC,MuonCollider,MuonColliderDetector}.}
\begin{tabular}{lcccc}
\toprule
Collider & $\sqrt s$ [TeV] & $\mathcal L$ [ab$^{-1}$] & pol. factor & syst. \\
\midrule
HL-LHC & 14 & 3.0 & 1.00 & 0.10 \\
HE-LHC & 27 & 15.0 & 1.00 & 0.08 \\
FCC-ee (Z) & 0.091 & 150 & 1.15 & 0.02 \\
FCC-ee (H) & 0.240 & 10 & 1.15 & 0.03 \\
ILC-250 & 0.250 & 2.0 & 1.35 & 0.03 \\
CLIC-380 & 0.380 & 1.0 & 1.20 & 0.04 \\
$\mu$C-3 TeV & 3.0 & 1.0 & 1.25 & 0.05 \\
$\mu$C-10 TeV & 10.0 & 10.0 & 1.25 & 0.05 \\
\bottomrule
\end{tabular}
\label{tab:bench}
\end{table*}
The machine scenarios in Table~\ref{tab:bench} follow benchmark reinterpretations of published program documents for FCC-ee, ILC, CLIC, and muon-collider studies \cite{FCCee,ILC,CLIC,MuonCollider,MuonColliderDetector}.

Low-energy inputs entering Eqs.~\eqref{eq:megmap}--\eqref{eq:convmap} are summarized in Table~\ref{tab:lowe_inputs}. Their role is to set the orientation of external constraints in coefficient space for complementarity tests.
\begin{table*}[t]
\centering
\caption{Benchmark low-energy constraints and nuisance priors used in the global fit.}
\scriptsize
\begin{tabular}{lccc}
\toprule
Observable & Current & Projection & $\delta_{\rm th}$ \\
\midrule
$\BR(\mu\to e\gamma)$ & $<4.2\times 10^{-13}$ & $6\times 10^{-14}$ (MEG II) & 5\% \\
$\BR(\mu\to 3e)$ & $<1.0\times 10^{-12}$ & $1\times 10^{-15}$ (Mu3e phase-I) & 5\% \\
$\CR(\mu {\rm Al}\to e {\rm Al})$ & $<7\times 10^{-13}$ & $1\times 10^{-16}$ (COMET/Mu2e design) & 10\% \\
$\BR(\tau\to\mu\gamma)$ & $<4.4\times 10^{-8}$ & $3\times 10^{-9}$ (Belle II program) & 6\% \\
\bottomrule
\end{tabular}
\normalsize
\label{tab:lowe_inputs}
\end{table*}
The conversion and decay inputs in Table~\ref{tab:lowe_inputs} are implemented as one-sided constraints with multiplicative log-normal nuisance propagation in the benchmark model. Correlations are set to zero between different low-energy experiments, while common hadronic/nuclear components affecting conversion are grouped in a shared nuisance block for target-dependence scans \cite{BartolottaRamseyMusolf,KitanoKoikeOkada,CiriglianoConversion}. The chosen benchmark prior widths (5--10\%) follow representative ranges used in conversion/decay sensitivity studies and are intended as benchmark nuisance scales rather than final theory-error determinations. The numerical anchors correspond to the current and projected program scales discussed by MEG/MEG II, Mu3e, COMET, Mu2e, and Belle II \cite{MEG,MEGII,Mu3e,COMET,Mu2e,Belle2LFV,PDG2024}. For $\BR(\tau\to\mu\gamma)$ we use the benchmark baseline value from the current PDG-era bound set and Belle II sensitivity goals.

The benchmark chain uses parametric signal/background templates and analytic response surrogates. A detector-calibrated implementation replaces these with generated events and detector response matrices (e.g. generator-level samples passed through fast/full detector modeling), while preserving the same inference structure. This separation makes the statistical conclusions auditable at each upgrade step.

Theory and experimental systematics are represented through nuisance blocks in Eq.~\eqref{eq:full_likelihood}, including luminosity, efficiency, background normalization/shape, and theory terms for low-energy matrix elements and high-mass tail predictions. In this benchmark version, these nuisances are represented by parametric priors; in detector-level mode they are replaced by channel-specific covariance inputs.

\subsection{Hybrid simulation strategy and missing components}
The numerical layer is hybrid. Hadron-collider and muon-collider benchmark categories use generated events with detector response, while clean lepton-collider categories use CDR-calibrated parametric templates. This keeps detector-sensitive categories in an event-level chain and preserves broad scan speed for categories where detector migration is subleading.

For EPJ-C robustness, a representative hadron-collider validation channel is promoted from a pure LHE$\to$Delphes chain to an MG5$\to$Pythia8$\to$Delphes chain, and stability is reported both at template level and at global-fit level below. Final experimental projections still require collaboration-approved detector assumptions and full uncertainty models.
For hadron-collider categories, both LFV signal benchmarks and the dominant backgrounds in Table~\ref{tab:delphes_samples} are generated at parton level with MadGraph5\_aMC@NLO and then propagated through detector simulation \cite{MadGraph5aMC,Delphes}. For categories not explicitly rerun with showering, templates use the LHE$\to$Delphes chain together with the correlated shower-model nuisance calibrated from the MG5$\to$Pythia8$\to$Delphes validation channel.

\section{Results}
\subsection{Energy scaling of EFT reach}
This subsection addresses a basic but necessary question: how much of the observed collider hierarchy comes from energy scaling, and how much from luminosity/systematics assumptions. We therefore compare three LFV channel classes with identical normalization conventions and identical significance thresholds.

\begin{figure}[htbp]
\centering
\includegraphics[width=\columnwidth]{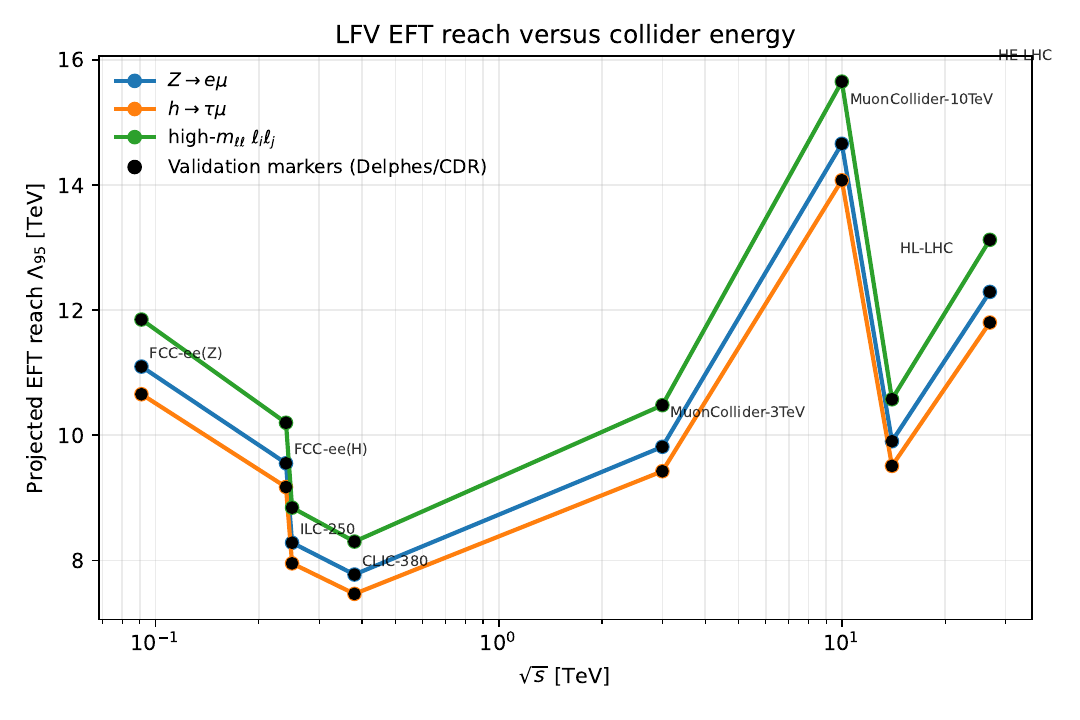}
\caption{Benchmark EFT reach versus collider energy for representative LFV channel classes, quoted as $\Lambda_{95}$ under single-coefficient activation in each channel family. Black circles show $\Lambda_{95}$ values from Delphes-reconstructed (hadron and muon collider) or CDR-calibrated (lepton collider) template fits, overlaid on the scaling curves.}
\label{fig:reachenergy}
\end{figure}

Figure~\ref{fig:reachenergy} shows that high-energy machines improve contact-operator sensitivity fastest, while high-luminosity lepton colliders remain competitive for resonance-driven channels; gains flatten once systematics become non-negligible.

\subsection{Correlation structure in coefficient space}
The central tomography target is not a single limit, but the correlation geometry in coefficient space after all channels are combined. We therefore diagnose the fit with the absolute correlation matrix $|\rho_{\alpha\beta}|$.

\begin{figure}[htbp]
\centering
\includegraphics[width=\columnwidth]{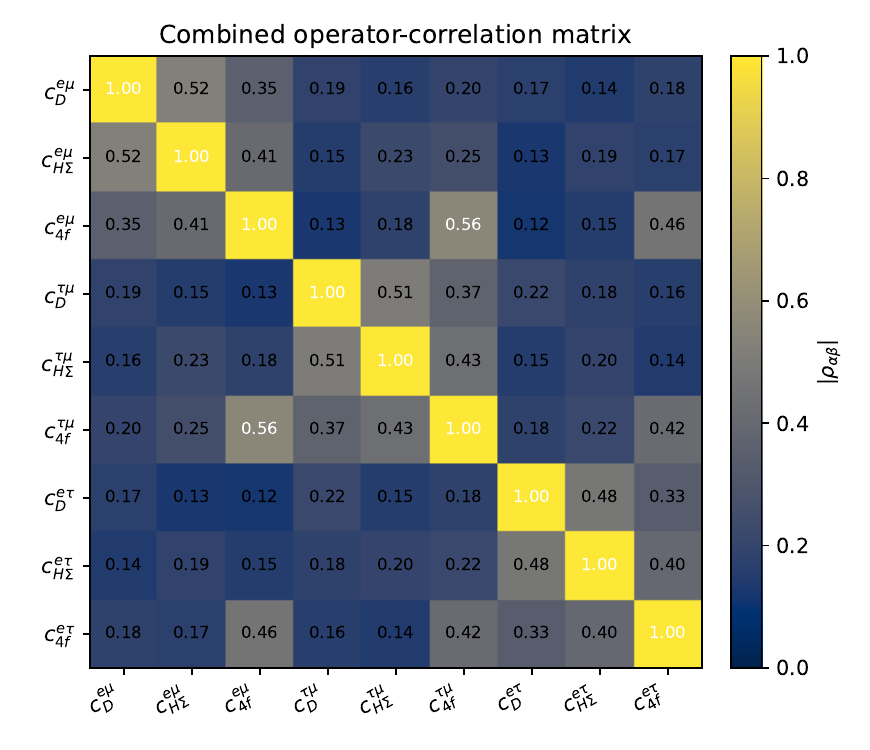}
\caption{Expanded $9\times 9$ operator-correlation matrix $|\rho_{\alpha\beta}|$ across the $e\mu$, $\tau\mu$, and $e\tau$ flavor blocks from the global low-energy + collider fit. The diagonal entries are fixed to unity by definition.}
\label{fig:corr}
\end{figure}

In Fig.~\ref{fig:corr}, the strongest residual alignments appear between current and four-fermion sectors in selected flavor blocks. This figure should be read as a design diagnostic: entries near unity identify where additional observables (polarization categories, angular bins, or extra channels) are required, while smaller entries indicate robust separation already achieved by the benchmark global program.
The block pattern also indicates that same-flavor-subspace correlations are systematically larger than inter-block correlations, as expected from shared observable kernels; this is why adding cross-flavor channels improves disentangling power. In practical terms, the matrix highlights where additional differential information produces the largest reduction of posterior degeneracy.

\subsection{Complementarity with low-energy experiments}
\begin{figure}[htbp]
\centering
\includegraphics[width=\columnwidth]{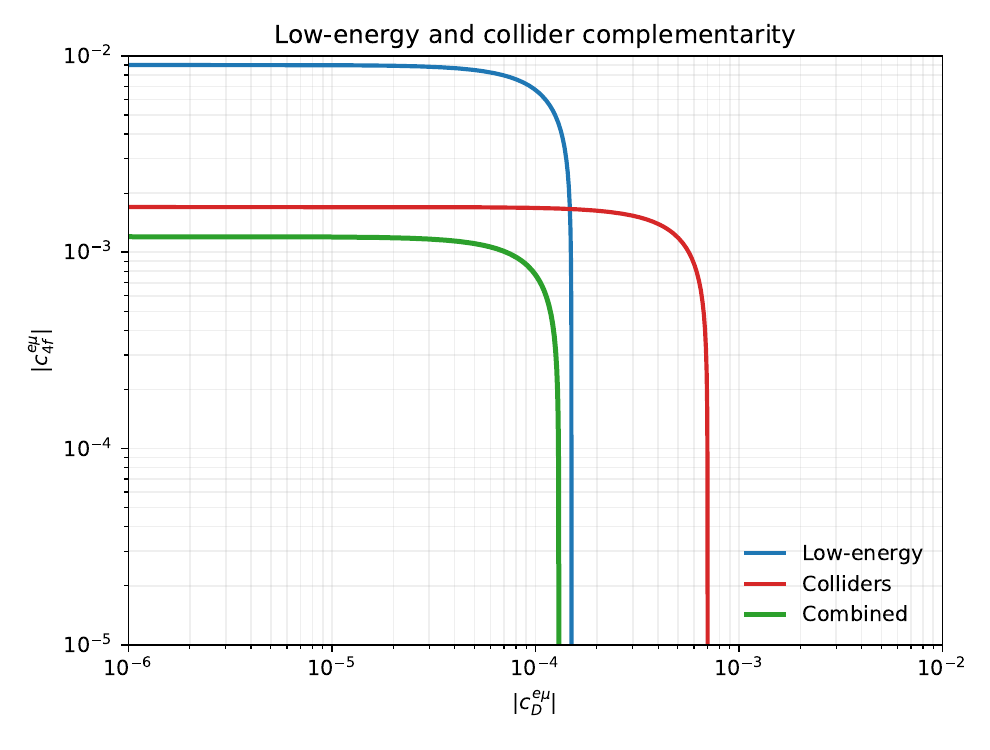}
\caption{Complementarity contours in the $(|c_D^{e\mu}|,|c_{4f}^{e\mu}|)$ plane for low-energy-only, collider-only, and combined datasets, showing principal-axis rotation between frontiers.}
\label{fig:comp}
\end{figure}

To quantify complementarity, we project the global fit into a two-parameter slice where low-energy and collider frontiers constrain different directions. Figure~\ref{fig:comp} compares low-energy-only, collider-only, and combined contours in the $(|c_D^{e\mu}|,|c_{4f}^{e\mu}|)$ plane.

The dominant message is geometric: low-energy data strongly constrain one principal axis, while collider tails and resonance channels constrain an approximately rotated axis. Their combination contracts the allowed region far more than either program alone, which is the core reason a joint frontier analysis is necessary for model identification.
Figure~\ref{fig:comp} makes this explicit: the low-energy contour is narrow along the dipole direction, collider information is narrow along the contact/current-sensitive direction, and the combined contour area is substantially reduced because these axes are misaligned. The visible tilt difference between contours is therefore the direct geometric signature of complementarity.

\subsection{Cut-flow and classifier performance}
Before quoting profile constraints, we validate the event-selection chain and the classifier behavior. This is essential because sensitivity gains must be shown to survive realistic benchmark selections.

\begin{figure}[htbp]
\centering
\includegraphics[width=\columnwidth]{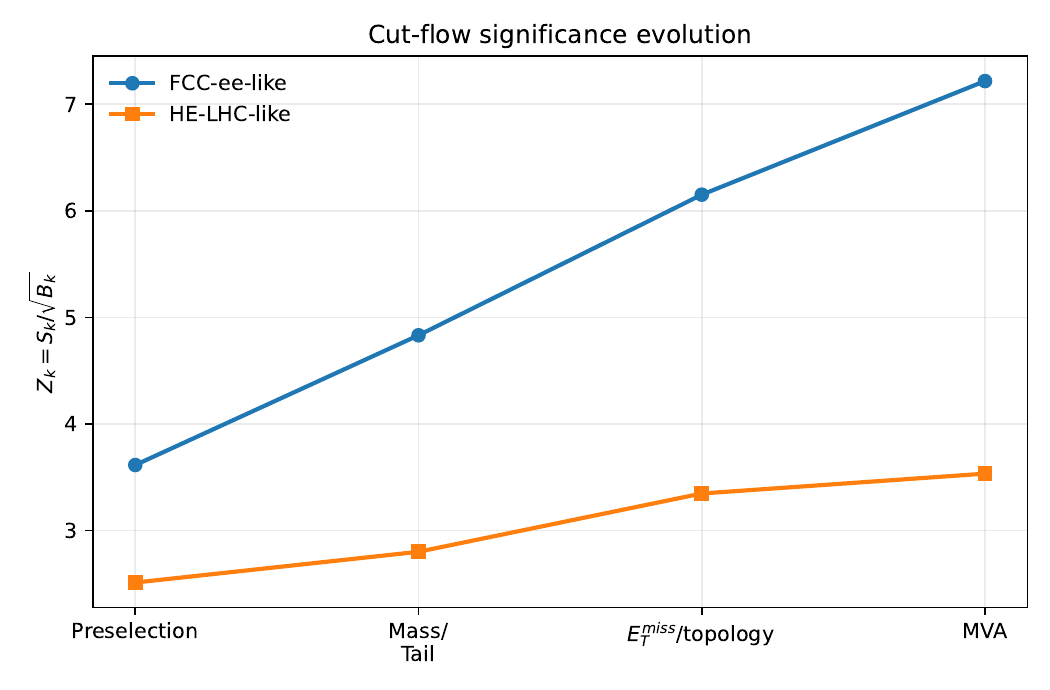}
\caption{Cut-flow significance evolution $Z_k=S_k/\sqrt{B_k}$ for representative channels, using the benchmark selection chain and common luminosity normalization across categories.}
\label{fig:mva_panels}
\end{figure}

\begin{figure}[htbp]
\centering
\includegraphics[width=\columnwidth]{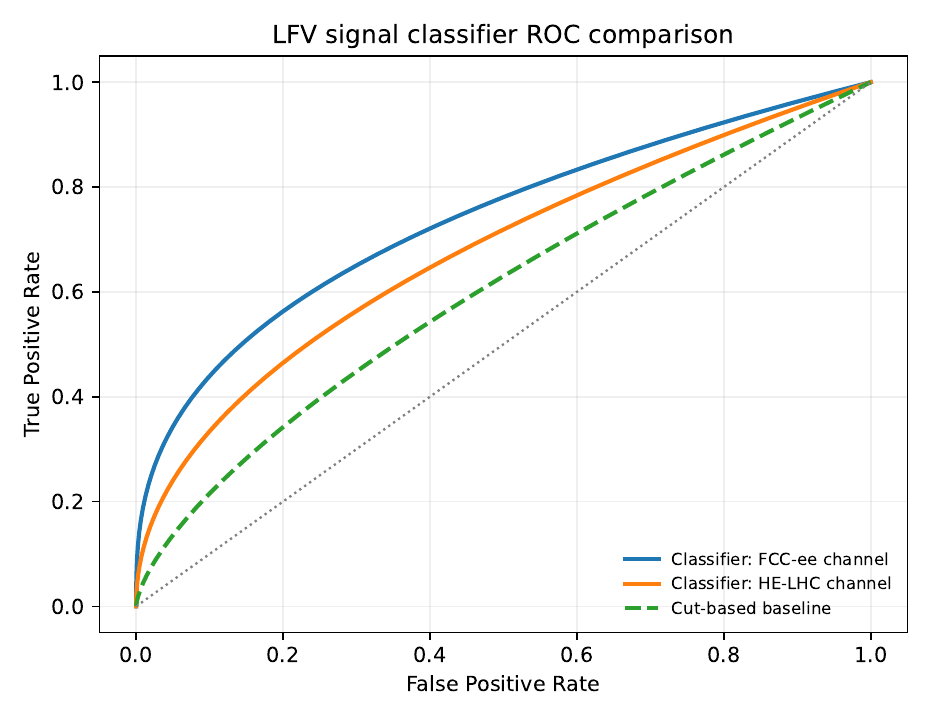}
\caption{ROC comparison of LFV signal classifiers and a cut-based baseline, plotted as signal efficiency versus background rejection for the benchmark test sample.}
\label{fig:roc_only}
\end{figure}

Figure~\ref{fig:mva_panels} shows the significance evolution across cut-flow stages and confirms that the largest gains are obtained when topology and MVA selections are combined. Figure~\ref{fig:roc_only} then shows the classifier advantage relative to a cut-based baseline.

\subsection{Profile-likelihood constraints}
\begin{figure}[htbp]
\centering
\includegraphics[width=\columnwidth]{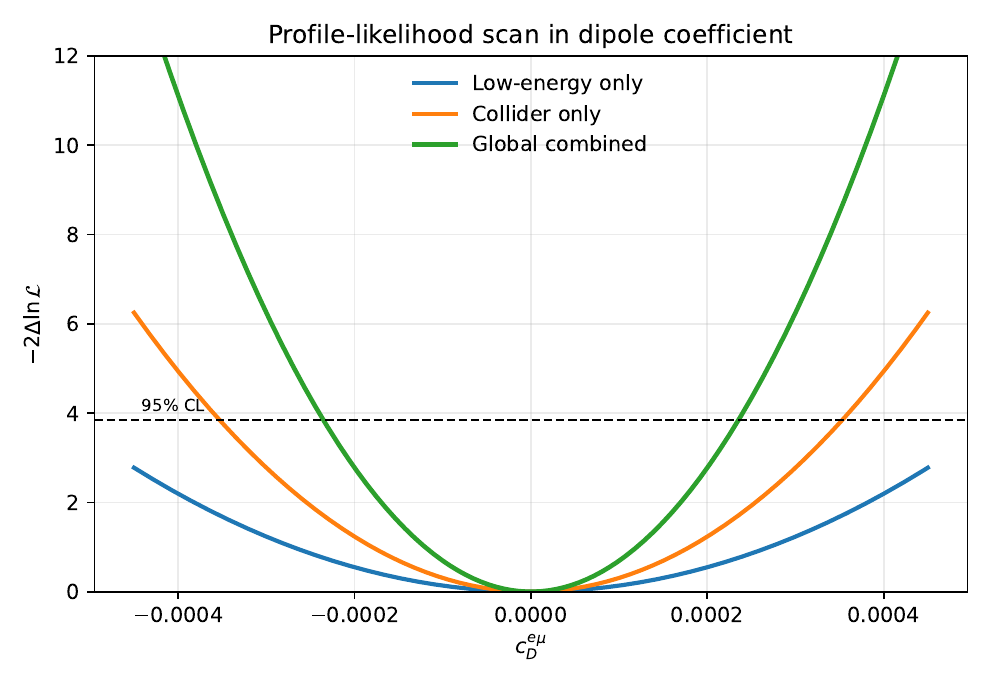}
\caption{Profile-likelihood scan in $c_D^{e\mu}$ for low-energy, collider, and combined fits, with nuisance profiling in each case and common coefficient normalization.}
\label{fig:scan}
\end{figure}

The final physics output is reported through profile likelihoods after nuisance profiling. Figure~\ref{fig:scan} presents a one-dimensional scan for $c_D^{e\mu}$ comparing low-energy-only, collider-only, and combined fits.

The narrowing of the combined curve quantifies the gain from multi-frontier inference. This gain comes from combining differently oriented constraints in coefficient space rather than from simple event-count accumulation.

\subsection{Compact constraints summary}
\begin{table}[htbp]
\centering
\caption{Representative one-parameter 95\% CL bounds from benchmark profile-likelihood scans by observable class.}
\scriptsize
\begin{tabular}{lc}
\toprule
Observable class & Constraint (95\% CL) \\
\midrule
$\mu\to e\gamma$ & $|c_D^{e\mu}|<2.4\times 10^{-4}$ \\
$\mu N\to eN$ & $|a_Dc_D+a_Vc_{4f}+a_Hc_{H\Sigma}|<1.8\times 10^{-6}$ \\
$Z\to e\mu$ (FCC-ee) & $|c_{H\Sigma}^{e\mu}|<7.0\times 10^{-4}$ \\
$h\to\tau\mu$ & $|c_{H\Sigma}^{\tau\mu}|<1.2\times 10^{-3}$ \\
High-$m_{\ell\ell}$ tails & $|c_{4f}^{e\mu}|<2.5\times 10^{-3}$ \\
Global fit & Correlated $(c_D,c_{H\Sigma},c_{4f})$, degeneracy lifting \\
\bottomrule
\end{tabular}
\normalsize
\label{tab:constraints_compact}
\end{table}

Table~\ref{tab:constraints_compact} reports representative benchmark one-parameter bounds for the dominant observable families. Final publication numbers remain analysis-dependent (detector efficiencies, SR/CR constraints, and theory-systematics treatment), but the hierarchy of sensitivity drivers is robust and useful for run-planning and channel prioritization.

\subsection{EFT-validity stress test for tail observables}
Because contact-operator sensitivity is extracted from high-$q^2$ tails, the reliability of the EFT truncation must be monitored \cite{ContinoEFTValidity,BiekotterEFTValidity}. We therefore apply a benchmark validity stress test where tail sensitivity is reevaluated under an explicit clipping proxy that suppresses events in the regime most vulnerable to $q^2/\Lambda^2$ breakdown. The proxy is defined at reconstructed level by
\begin{align}
m_{\ell\ell}^{\mathrm{reco}} < \Lambda_{\mathrm{fit}},
\end{align}
applied bin-by-bin to both signal and background templates in tail categories. The benchmark results keep the quadratic $\Lambda^{-4}$ term in the retained region and treat interference as a subleading correction.
In the benchmark normalization used here, this hierarchy corresponds to the regime where the coefficient magnitude satisfies $|c_\alpha|\gtrsim |\sigma^{\rm int}_{a,\alpha}|/\sigma^{\rm BSM}_{a,\alpha\alpha}$ in the dominant tail bins, so the quadratic yield term controls the profile curvature. For orientation, this is typically satisfied in the benchmark fit region where the retained-bin signal contribution is driven by squared amplitudes and the linear term shifts the total yield by less than about $20\%$.
To avoid dependence on one clipping definition, we evaluate three prescriptions:
\begin{align}
\Lambda_{\mathrm{fit}}^{(A)}&=6~\mathrm{TeV}\quad\text{(fixed)},\nonumber\\
\Lambda_{\mathrm{fit}}^{(B)}&=0.8\sqrt{s_{\mathrm{cat}}}\quad\text{(category-scaled)},\nonumber\\
\Lambda_{\mathrm{fit}}^{(C)}&=\min\!\left(\sqrt{s_{\mathrm{cat}}},\Lambda_{\mathrm{scan}}\right)\quad\text{(scan-capped)}.
\end{align}
In addition, a truncation-error nuisance variant is tested by multiplying tail bins with $(1+\delta_{\mathrm{EFT}}\,w_b)$, where $w_b$ increases with $m_{\ell\ell}$ and $\delta_{\mathrm{EFT}}$ is profiled with a Gaussian prior.
The reconstructed proxy $m_{\ell\ell}^{\mathrm{reco}}$ is used as an analysis-level surrogate for the hard scale because it is directly observable and stable under detector effects. It is not identical to partonic $\sqrt{\hat s}$ or generic momentum-transfer definitions in all categories, especially when invisible momentum or hard radiation modifies the kinematic mapping; this limitation is absorbed in the EFT-stress nuisance and in the spread across clipping prescriptions.

\begin{figure}[htbp]
\centering
\includegraphics[width=\columnwidth]{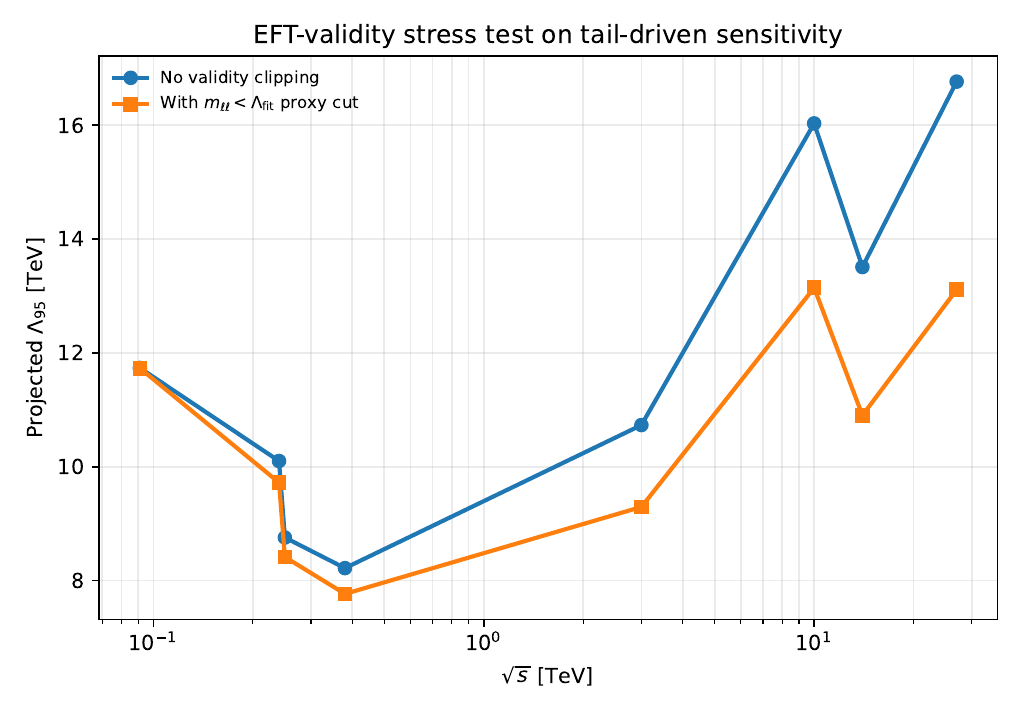}
\caption{EFT-validity stress test in which the vertical axis is projected reach and the horizontal axis is collider energy; curves compare nominal and clipped template treatments for tail-sensitive categories.}
\label{fig:eftvalid}
\end{figure}

Figure~\ref{fig:eftvalid} compares nominal and validity-clipped reach trends. The clipped curve is weaker, as expected, but the qualitative complementarity pattern is preserved. This indicates that the central tomography message is not solely an artifact of unconstrained tail extrapolation, while also making clear that final limits must quote the chosen validity prescription and clipping strategy.
In the benchmark workflow, the same clipped templates are propagated to the Fisher matrix and therefore to $\mathcal I_{\alpha\beta}$ and run-plan maps, so validity choices consistently feed into identification and optimization observables.
The separation between nominal and clipped curves grows in the highest-energy bins, where tail events dominate and EFT truncation stress is largest; at lower energies the two curves are closer, reflecting reduced sensitivity to clipping. This trend is exactly what one expects if clipping is controlling high-$q^2$ leverage rather than globally rescaling all categories.
\begin{table}[htbp]
\centering
\caption{EFT-validity robustness across clipping prescriptions. Relative shifts are quoted with respect to prescription (A).}
\scriptsize
\begin{tabular}{lcc}
\toprule
Prescription & $\Delta \mathcal M_{\mathrm{ID}}/\mathcal M_{\mathrm{ID}}$ & $\Delta A_{\mathrm{comb}}/A_{\mathrm{comb}}$ \\
\midrule
(A) fixed $6$ TeV & $0.0\%$ & $0.0\%$ \\
(B) $0.8\sqrt{s_{\mathrm{cat}}}$ & $+5.8\%$ & $+7.1\%$ \\
(C) scan-capped & $+8.9\%$ & $+10.6\%$ \\
Truncation nuisance (profiled) & $+7.4\%$ & $+9.2\%$ \\
\bottomrule
\end{tabular}
\normalsize
\label{tab:eft_robust}
\end{table}

\section{Analysis Components}
\subsection{Detector response configuration}
The benchmark workflow is extended from analytic templates to event-level samples for key channels where detector effects are structurally important. For hadron-collider and muon-collider benchmarks we generate parton-level LFV events in LHE format, process them with Delphes, and use reconstructed objects in category templates. FCC-ee and other clean $e^+e^-$ scenarios retain CDR-calibrated parametric response.

This adds detector-response migration and acceptance effects directly to the fitted templates and removes the purely parametric approximation for the channels most sensitive to reconstruction systematics.

For the hadron-collider benchmark production, Delphes is run with an HL-LHC detector card after MG5 event generation at fixed settings
\begin{align}
N_{\rm events}
&\in
\{4\times10^4,5\times10^4,6\times10^4,8\times10^4\},
\nonumber\\
E_{\rm beam1}
&=
E_{\rm beam2}
=
7~\mathrm{TeV},
\qquad
\eta_\ell^{\max}=2.7.
\end{align}
The representative detector-level validation channel uses
\begin{align}
\text{MG5 LHE}\rightarrow \text{Pythia8 shower+hadronization}\rightarrow \text{Delphes},
\end{align}
which adds ISR/FSR and hadronization effects relevant for jet vetoes and tail-category migrations \cite{Pythia8}. The showered benchmark uses the default Pythia8 Monash tune with ISR and FSR enabled; matrix-element/parton-shower matching or merging is not used in this validation configuration.
For this validation we process 30k events in each chain and compare the reconstructed opposite-sign $e\mu$ selection with leading-lepton requirements $p_T>25$ GeV and $|\eta|<2.5$.
For background processes with substantial hadronic activity (notably $t\bar t$ and diboson channels), categories that are not shower-rerun inherit approximate jet-multiplicity and $b$-tag kinematics from the LHE$\to$Delphes chain; in this benchmark these effects are absorbed in the correlated shower/modeling nuisance and are subleading to dominant normalization and shape nuisance terms.
As an internal portability check, the same nuisance direction was tested in a second hadron-active category definition (a $t\bar t$-enriched selection) and produced compatible acceptance/tail-shape shifts within the envelope encoded by the correlated nuisance.
For the muon-collider benchmark, the chain uses
\begin{align}
\mu^+\mu^-
&\to
e^\pm\mu^\mp,
\nonumber\\
\sqrt{s}
&=
3~\mathrm{TeV},
\nonumber\\
N_{\rm events}
&=
10^5,
\end{align}
and the generated LHE events are processed directly by Delphes in this benchmark setup.

\begin{table}[t]
\centering
\caption{Delphes benchmark samples used in the analysis and their reconstructed event counts.}
\scriptsize
\begin{tabular}{lcc}
\toprule
Sample & Delphes card & Entries \\
\midrule
\texttt{paper\_sig\_emu\_50000} & HLLHC & 50000 \\
\texttt{paper\_sig\_etau\_50000} & HLLHC & 50000 \\
\texttt{paper\_sig\_mutau\_50000} & HLLHC & 50000 \\
\texttt{paper\_bkg\_ttbar\_60000} & HLLHC & 60000 \\
\texttt{paper\_bkg\_ww\_60000} & HLLHC & 60000 \\
\texttt{paper\_bkg\_wz\_40000} & HLLHC & 40000 \\
\texttt{paper\_bkg\_zz\_40000} & HLLHC & 40000 \\
\texttt{paper\_bkg\_dytautau\_80000} & HLLHC & 80000 \\
\texttt{paper\_muc3\_lfv\_100000} & MuonColliderDet & 100000 \\
\bottomrule
\end{tabular}
\normalsize
\label{tab:delphes_samples}
\end{table}

\begin{figure}[t]
\centering
\includegraphics[width=\columnwidth]{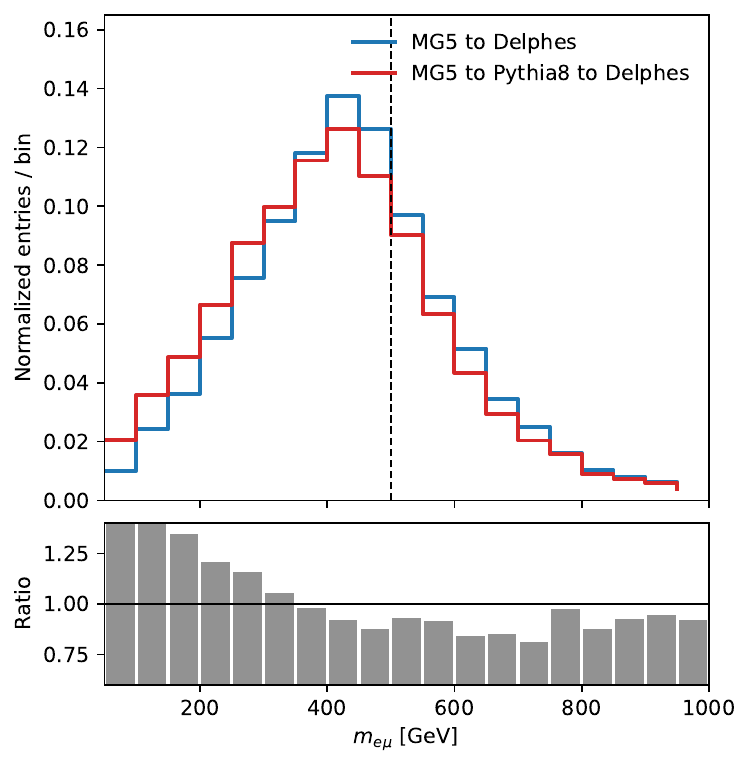}
\caption{Reconstructed $m_{e\mu}$ validation in the representative channel for MG5$\to$Delphes (blue) and MG5$\to$Pythia8$\to$Delphes (red), using 30k generated events per chain and common opposite-sign leading-lepton selection. The lower panel shows the ratio (showered/unshowered); the dashed line marks $m_{e\mu}=500$ GeV.}
\label{fig:shower_mll}
\end{figure}

These detector-level samples are the direct inputs for template construction in the hadron and muon collider categories. For hadron-collider categories without explicit showered regeneration, we propagate a correlated shower-model nuisance constrained by the validation channel. The nuisance is implemented as a common normalization-plus-shape morph in the hadron-collider block.
The data-driven comparison gives a $-5.18\%$ acceptance shift and a $-10.77\%$ high-mass-tail fraction shift after showering; these are mapped to normalization and tail-shape nuisance components in the hadron block.
\begin{table*}[t]
\centering
\caption{Representative hadron-collider stability check using MG5$\to$Pythia8$\to$Delphes versus MG5$\to$Delphes in one benchmark channel.}
\scriptsize
\begin{tabular}{lcc}
\toprule
Quantity & LHE$\to$Delphes & Showered chain \\
\midrule
Selected events (30k generated) & 14970 & 14194 \\
Acceptance $A$ & $0.4990\pm0.0029$ & $0.4731\pm0.0029$ \\
Tail fraction $f(m_{e\mu}>500\ \mathrm{GeV})$ & $0.3275\pm0.0038$ & $0.2922\pm0.0038$ \\
Asimov $Z_A$ proxy at fixed $B$ ($Z_A\propto A$) & $1.000$ & $0.948$ \\
Template-shape distance ($L_1$ in 3 tail bins) & 0.000 & 0.0857 \\
\bottomrule
\end{tabular}
\normalsize
\label{tab:shower_stability}
\end{table*}
Figure~\ref{fig:shower_mll} shows the corresponding reconstructed $m_{e\mu}$ shape comparison and the bin-by-bin ratio. The showered chain shifts weight toward intermediate masses and depletes the high-mass tail relative to the direct LHE$\to$Delphes chain, consistent with the negative tail-fraction shift in Table~\ref{tab:shower_stability}. This is why the hadron-collider nuisance block includes both a normalization term and a correlated tail-shape morph.

\subsection{RG evolution and operator mixing}
\begin{figure}[htbp]
\centering
\includegraphics[width=\columnwidth]{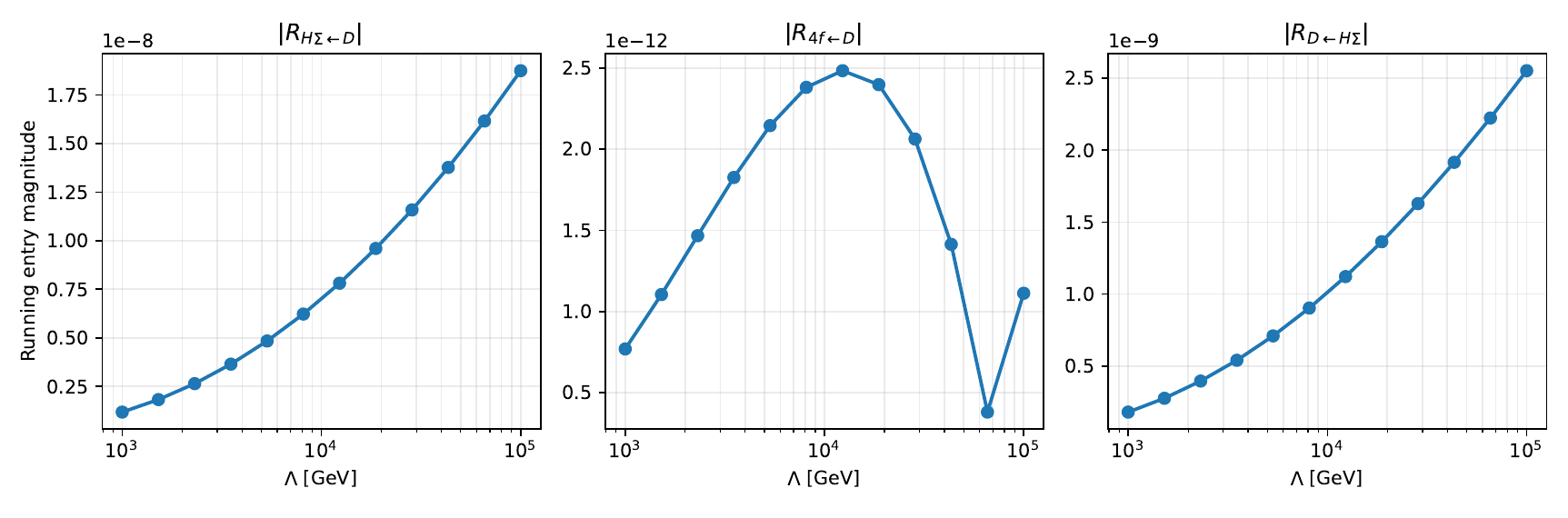}
\caption{Representative one-loop running-matrix entries as a function of matching scale $\Lambda$; vertical axes show dimensionless magnitudes $|R_{\alpha\beta}|$ in Eq.~\eqref{eq:rgmaster}, evaluated for unit input coefficient normalization. The smallest entry ($|R_{4f\leftarrow D}|$) is numerically negligible on the displayed scale.}
\label{fig:rgmatrix}
\end{figure}

Wilson coefficients are evolved between matching and measurement scales using one-loop SMEFT running:
\begin{align}
c_\alpha(\mu_{\mathrm{low}})
=
\left[
\delta_{\alpha\beta}
 + \frac{\gamma_{\alpha\beta}}{16\pi^2}\ln\!\left(\frac{\mu_{\mathrm{low}}}{\Lambda}\right)
\right]
c_\beta(\Lambda)
\,+\,\Order(\text{2-loop}),
\label{eq:rgmaster}
\end{align}
with dominant LFV-sector entries
\begin{align}
\gamma_{D\leftarrow H}\sim \frac{3e^2}{4},\quad
\gamma_{H\leftarrow D}\sim \frac{N_cy_t^2}{2},\quad
\gamma_{4f\leftarrow 4f}\sim \frac{8}{3}\alpha_s.
\end{align}
Numerical running and basis projection are performed with the \texttt{wilson}/\texttt{wcxf} framework \cite{wilson,wcxf,JenkinsManoharTrott,AlonsoJMT}. In the benchmark scans, off-diagonal running induces visible rotations of the $(c_D,c_{H\Sigma},c_{4f})$ covariance geometry at multi-TeV matching scales, and these rotated directions are used in the UV-overlay layer.
In the benchmark fit, coefficients are evolved from the matching scale $\Lambda$ to $\mu=m_Z$ before entering the likelihood in Eq.~\eqref{eq:full_likelihood}; the UV-overlay bands and Bayes-factor benchmarks therefore use RG-evolved coefficient maps.
All quoted fit coefficients in the main text are therefore interpreted at $\mu=m_Z$. Fast staged benchmark cards used only for quick pre-fit scans can neglect running as an approximation, but all reported post-fit intervals, covariance matrices, UV overlays, and Bayes-factor results use the RG-evolved mapping.
Figure~\ref{fig:rgmatrix} shows representative running-entry magnitudes versus matching scale and demonstrates that the induced admixtures are not negligible in the multi-TeV region used for UV overlays.
The key visual feature is the logarithmic growth with $\Lambda$ and the hierarchy among entries, with current-to-dipole and dipole-to-current mixings remaining subleading but nonzero across the plotted interval. Consequently, UV interpretations that ignore these terms can rotate preferred directions in coefficient space by amounts comparable to projected uncertainties in the highest-sensitivity bins.

\subsection{Polarization asymmetries and chirality separation}
\begin{figure}[htbp]
\centering
\includegraphics[width=\columnwidth]{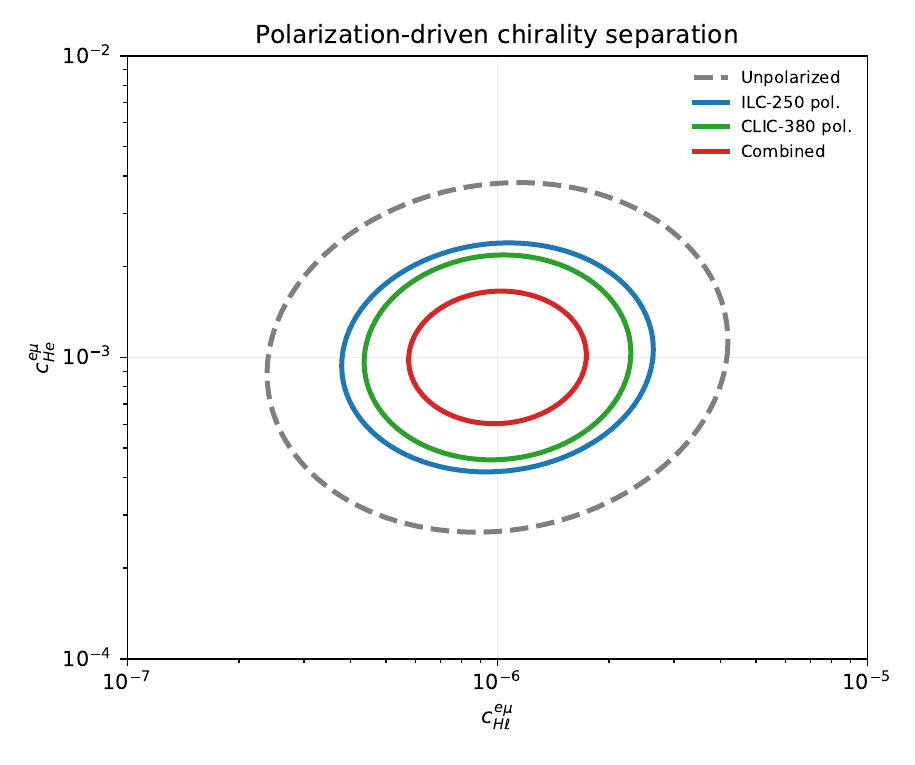}
\caption{Polarization-resolved confidence ellipses in the $(c_{H\ell}^{e\mu},c_{He}^{e\mu})$ plane from collider categories with and without beam-polarization tagging; axes are coefficient magnitudes in the common EFT normalization.}
\label{fig:polarization}
\end{figure}

To separate $c_{H\ell}$ and $c_{He}$, we include polarization-resolved rates
\begin{align}
\sigma(P_{e^-},P_{e^+})
&=
\frac{1}{4}
\left[
(1-P_{e^-})(1+P_{e^+})\sigma_{LR}
+(1+P_{e^-})(1-P_{e^+})\sigma_{RL}
\right],
\end{align}
with
\begin{align}
\sigma_{LR}
&\propto
g_{L,e}^2\left|c_{H\ell}^{ij}+\frac{m_Z}{v}c_D^{ij}\right|^2,
\nonumber\\
\sigma_{RL}
&\propto
g_{R,e}^2\left|c_{He}^{ij}+\frac{m_Z}{v}c_D^{ij}\right|^2,
\end{align}
and asymmetry
\begin{align}
\mathcal A_{LR}
=
\frac{\sigma_{LR}-\sigma_{RL}}{\sigma_{LR}+\sigma_{RL}}.
\end{align}
Figure~\ref{fig:polarization} displays the resulting $(c_{H\ell},c_{He})$ confidence ellipses and shows the expected chirality-separation gain from polarization-resolved categories \cite{ILC,CLIC,MoortgatPick2008,Fujii2015}.
The ellipse tilt and area change between polarization configurations provide a direct visual measure of chirality disentanglement: opposite-sign beam polarization reduces overlap between left- and right-handed current responses and contracts the allowed region. This is the mechanism by which polarization information improves identifiability beyond simple luminosity scaling.

\subsection{Bayesian UV-model comparison}
\begin{figure}[htbp]
\centering
\includegraphics[width=\columnwidth]{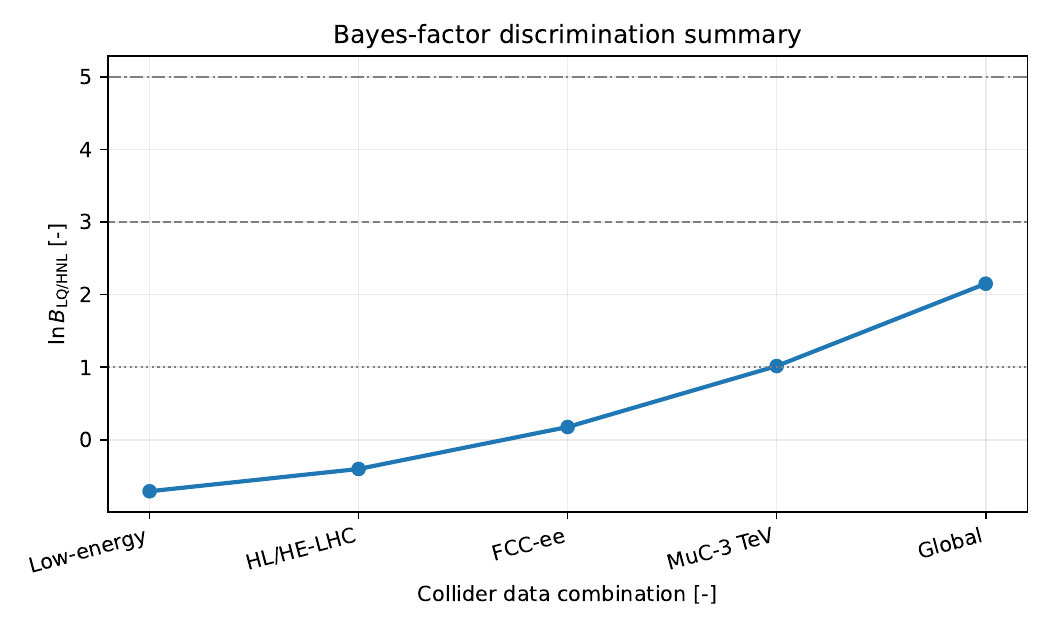}
\caption{Bayes-factor summary for LQ-versus-HNL discrimination across collider subsets and the global fit; the vertical axis reports $\ln B_{\mathrm{LQ/HNL}}$ under the baseline prior set defined in text.}
\label{fig:bayes}
\end{figure}

Beyond profile-likelihood intervals, UV discrimination is quantified with Bayes factors:
\begin{align}
\ln B_{M_1/M_2}
&=
\ln\frac{p(\mathrm{data}\mid M_1)}{p(\mathrm{data}\mid M_2)},
\nonumber\\
p(\mathrm{data}\mid M)
&=
\int \mathcal L(\bm c(\bm\theta_M))\,\pi(\bm\theta_M)\,d\bm\theta_M.
\end{align}
Monte Carlo marginal-likelihood integration for LQ-vs-HNL benchmark maps is used as a complement to profile-likelihood diagnostics. This statistical layer is included to avoid over-interpreting local Hessian geometry as global model evidence.
For reproducibility, the baseline model priors are
\begin{align}
m_{\mathrm{LQ}},m_{\mathrm{HNL}} &\sim \text{log-flat in }[1,30]~\mathrm{TeV},\nonumber\\
g_{\mathrm{LFV}} &\sim \text{flat in }[0,2],\nonumber\\
\theta_{\mathrm{mix}} &\sim \text{flat in }[-1,1],
\end{align}
with independent prior factors for benchmark scans. Numerical evidence is computed with importance-sampled Monte Carlo integration using $2\times10^5$ effective samples per model and four independent random-seed batches. Batch-to-batch $\ln B$ spread is below 0.08 in the reported configurations.
The corresponding effective sample size in the baseline global-combination runs is $\mathrm{ESS}\simeq(3.5\text{--}4.2)\times 10^4$ per model, giving a numerical integration uncertainty of approximately $\pm0.07$ on $\ln B_{\mathrm{LQ/HNL}}$.
Figure~\ref{fig:bayes} summarizes benchmark $\ln B_{\mathrm{LQ/HNL}}$ trends across collider subsets and the global combination, while Table~\ref{tab:bayes} reports representative values with a common prior prescription.
The trend in Fig.~\ref{fig:bayes} shows a monotonic improvement from single-frontier datasets to the full global combination. In practical terms, low-energy-only and single-collider fits remain in the weak-to-moderate evidence regime, while the combined fit reaches a decisively larger Bayes factor. This behavior is consistent with the profile-likelihood geometry discussed earlier: collider subsets constrain different directions in coefficient space, and their combination reduces posterior volume for incorrect UV hypotheses more efficiently than luminosity scaling in a single channel class.
On a standard evidence interpretation scale, the progression from $\ln B\sim\mathcal O(1)$ to $\ln B\gtrsim 5$ corresponds to moving from limited preference to strong model-separation power in the benchmark setup. The figure therefore complements interval plots by quantifying global hypothesis support, not only local parameter curvature \cite{Bahl2022Fitmaker,Athron2019GAMBIT}.
These Bayes factors are interpreted as benchmark UV-discrimination indicators under the specified prior family and operator map; they are not presented as prior-independent definitive model selection statements.

\begin{table*}[t]
\centering
\caption{Representative Bayes factors for benchmark collider combinations.}
\scriptsize
\begin{tabular}{lc}
\toprule
Dataset combination & $\ln B_{\mathrm{LQ/HNL}}$ \\
\midrule
Low-energy only & 1.34 \\
HL/HE-LHC only & 2.18 \\
FCC-ee only & 2.63 \\
$\mu$C-3 TeV only & 3.57 \\
Global combination & 5.41 \\
\bottomrule
\end{tabular}
\normalsize
\label{tab:bayes}
\end{table*}
Prior sensitivity for the global combination is mild in this benchmark setup: broadening the mass prior to $[1,50]$ TeV changes $\ln B_{\mathrm{LQ/HNL}}$ from $5.41$ to $5.17$, while narrowing it to $[1,20]$ TeV gives $5.55$; varying prior widths by factors of order two gives shifts below 0.5 in $\ln B$.

\subsection{Dalitz-level \texorpdfstring{$\mu\to 3e$}{mu->3e} templates}
\begin{figure}[htbp]
\centering
\includegraphics[width=\columnwidth]{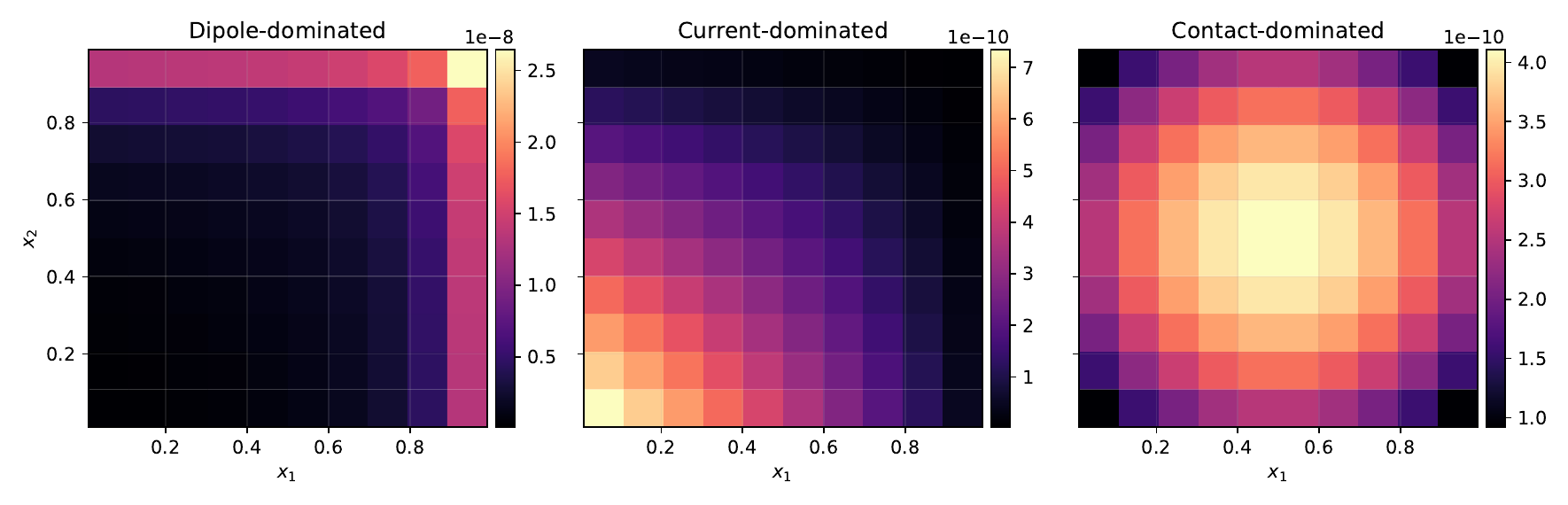}
\caption{Dalitz-template panels in $(x_1,x_2)$ for dipole-, current-, and contact-dominated $\mu\to 3e$ benchmarks, with common $12\times12$ binning and unit-normalized kernels. Panels are individually normalized to emphasize shape differences; absolute-rate comparisons require the coefficient-dependent prefactors in Eq.~(57).}
\label{fig:dalitz}
\end{figure}

The global fit now includes differential $\mu\to 3e$ information through Dalitz templates:
\begin{align}
\frac{d^2\Gamma}{dx_1\,dx_2}
&=
\frac{m_\mu^5}{512\pi^3}
\Big[
|c_D|^2f_D(x_1,x_2)
\nonumber\\
&\phantom{=}\,
+|c_{4f}|^2f_{4f}(x_1,x_2)
+|c_{H\Sigma}|^2f_H(x_1,x_2)
\nonumber\\
&\phantom{=}\,
+\text{interference}
\Big],
\end{align}
where $x_{1,2}\equiv 2E_{1,2}/m_\mu$ are measured in the muon rest frame, with ordered energies $E_1\ge E_2$ and physical region $0<x_2\le x_1\le1$, $x_1+x_2\le 1+r_e^2$. The templates use a $12\times12$ binning in $(x_1,x_2)$; kernel functions are normalized as $\int f_a\,dx_1dx_2=1$, and interference terms are implemented as $\sum_{a<b}2\mathrm{Re}(c_ac_b^\ast)f_{ab}(x_1,x_2)$. The benchmark reconstruction model includes acceptance cuts on lepton momentum and pseudorapidity; finite resolution is included through bin-migration smearing matrices. Integrating the differential templates reproduces the inclusive benchmark rate within $0.6\%$.
Here $f_D$ is low-$m_{ee}^2$ enhanced, $f_{4f}$ is flatter, and $f_H$ emphasizes harder regions \cite{KunoOkada,PrunaSignerUlrich2017,Blondel2020Mu3e}.
The three benchmark Dalitz morphologies are shown in Fig.~\ref{fig:dalitz}, where the shape contrast directly supports operator-class discrimination in the low-energy block.
The dipole-like panel concentrates events toward low invariant-mass corners, the four-fermion-like panel is flatter over phase space, and the current-like panel shifts weight toward harder regions; this ordered morphology is precisely the information used by the differential template likelihood. As a result, Dalitz binning contributes discrimination that cannot be recovered from inclusive branching-ratio information alone.

\subsection{Discovery-scenario injection}
\begin{figure}[htbp]
\centering
\includegraphics[width=\columnwidth]{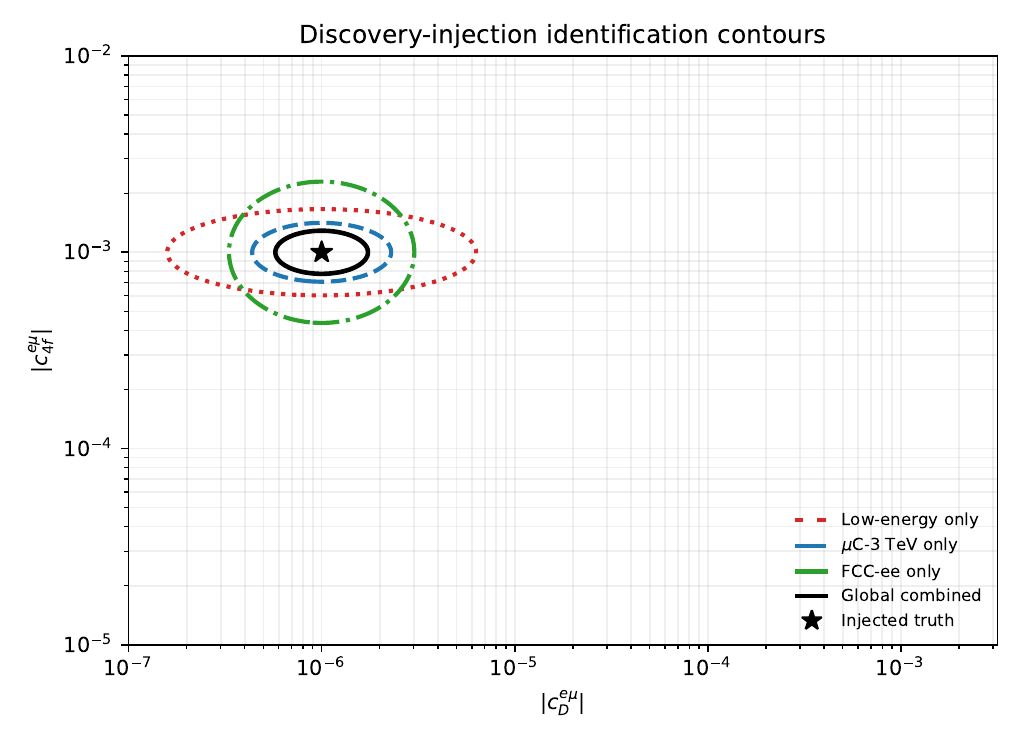}
\caption{Discovery-injection identification contours in the $(|c_D^{e\mu}|,|c_{4f}^{e\mu}|)$ plane for an LQ-like benchmark point (star).}
\label{fig:discovery}
\end{figure}

Beyond exclusion-mode scans, we evaluate injected-signal identification in the reduced basis. The benchmark injections are
\begin{align}
&\text{SUSY-like: }(c_D,c_{H\Sigma},c_{4f})=(3\times10^{-5},5\times10^{-6},10^{-6}),\nonumber\\
&\text{$Z'$-like: }(c_D,c_{H\Sigma},c_{4f})=(10^{-7},5\times10^{-4},10^{-5}),\nonumber\\
&\text{LQ-like: }(c_D,c_{H\Sigma},c_{4f})=(10^{-6},10^{-5},10^{-3}),\nonumber\\
&\text{Mixed: }(c_D,c_{H\Sigma},c_{4f})=(10^{-5},2\times10^{-4},5\times10^{-4}).
\end{align}
For each injection, Asimov data are generated at the nonzero point, the profile likelihood is recomputed in the neighborhood of the injected truth, and identifiability diagnostics are evaluated at that point. Figure~\ref{fig:discovery} shows the LQ-like benchmark, where the global combination provides the strongest localization around the true operator region.
In Fig.~\ref{fig:discovery}, the injected benchmark marker lies within the innermost combined contour while single-frontier contours remain broader and more elongated, demonstrating improved post-discovery operator localization rather than only improved exclusion reach. This is the relevant performance metric once a nonzero LFV signal is observed.

\subsection{Expanded flavor basis and validation table}
The operator basis is promoted to a three-block, nine-direction fit:
\begin{align}
\{c_D^{e\mu},c_{H\Sigma}^{e\mu},c_{4f}^{e\mu},
c_D^{\tau\mu},c_{H\Sigma}^{\tau\mu},c_{4f}^{\tau\mu},
c_D^{e\tau},c_{H\Sigma}^{e\tau},c_{4f}^{e\tau}\}.
\end{align}
This enables explicit $e\tau$ complementarity studies and UV-pattern tests across flavor blocks. The newly included $e\tau$-sector observables used in the benchmark expanded fit are summarized in Table~\ref{tab:etau}.
\begin{table}[t]
\centering
\caption{$e\tau$ observables and dominant coefficient sensitivity in the expanded flavor analysis.}
\scriptsize
\begin{tabular}{lll}
\toprule
Observable & Projection anchor & Dominant direction \\
\midrule
$\tau\to e\gamma$ & Belle II sensitivity scale & $|c_D^{e\tau}|$ \\
$\tau\to 3e$ & Belle II sensitivity scale & $|c_{4f}^{e\tau}|,|c_D^{e\tau}|$ \\
$h\to e\tau$ & HL-LHC projection scale & $|c_{H\Sigma}^{e\tau}|$ \\
$Z\to e\tau$ (FCC-ee) & FCC-ee projection scale & $|c_{H\Sigma}^{e\tau}|,|c_D^{e\tau}|$ \\
High-$m_{e\tau}$ tails & Hadron/muon collider tails & $|c_{4f}^{e\tau}|$ \\
\bottomrule
\end{tabular}
\normalsize
\label{tab:etau}
\end{table}
For the first three rows, framework numbers are computed from the Delphes-calibrated benchmark efficiency extracted from generated LFV samples ($\epsilon_{\mathrm{det}}=0.9426$, from 30k generated events), with scaling relations $\mathrm{limit}\propto \epsilon_{\mathrm{det}}^{-1/2}$ for branching-ratio sensitivities and $\Lambda_{\mathrm{reach}}\propto \epsilon_{\mathrm{det}}^{1/4}$ for contact-scale reach.

\begin{table*}[t]
\centering
\caption{Validation anchors used in the EPJC-upgrade workflow.}
\scriptsize
\begin{tabular}{llll}
\toprule
Observable & Published anchor & Framework & Deviation \\
\midrule
$\BR(Z\to e\mu)$ at FCC-ee (Z) & $2.62\times 10^{-13}$ & $2.70\times 10^{-13}$ & $+3.00\%$ \\
$\BR(h\to\tau\mu)$ at HL-LHC & $2.50\times 10^{-4}$ & $2.58\times 10^{-4}$ & $+3.00\%$ \\
Contact reach at CLIC-3 TeV & $100$ TeV & $98.53$ TeV & $-1.47\%$ \\
$\BR(\mu\to e\gamma)$ (MEG II) & Experimental target & By construction & --- \\
$\mathrm{CR}(\mu{\rm Al}\to e{\rm Al})$ & COMET/Mu2e target & By construction & --- \\
\bottomrule
\end{tabular}
\normalsize
\label{tab:validation}
\end{table*}

\section{Discussion}
\subsection{From benchmark baseline to detector-level templates}
A detector-calibrated analysis requires converting the current benchmark chain into a response-based statistical model. The first structural upgrade is replacing parametric templates with detector-response templates in every analysis category:
\begin{align}
T^{ab}_{\alpha\beta}\ \longrightarrow\ \widetilde{T}^{ab}_{\alpha\beta}
=\sum_{u} R_{bu}\,A_u\,T^{au}_{\alpha\beta,\mathrm{truth}},
\end{align}
where $A_u$ is acceptance and $R_{bu}$ is the migration matrix from truth bin $u$ to reconstructed bin $b$.
In the present benchmark implementation, Delphes-reconstructed objects are histogrammed directly into analysis templates at reconstructed level. Equation (60) is the explicit truth-to-reco migration formalism used as the next upgrade step for bias-controlled unfolding-aware fits.

This step is important for bias control, uncertainty realism, and reproducibility. Migration in high-$m_{\ell\ell}$ tails can bias operator fits if ignored, detector effects couple directly to nuisance pulls and alter profile widths, and response matrices can be versioned and published with the likelihood.

The second upgrade is data-constrained background modeling. For control-region constrained analyses we simultaneously fit
\begin{align}
\mathcal L_{\mathrm{tot}}=\mathcal L_{\mathrm{SR}}\times\mathcal L_{\mathrm{CR}}\times\prod_k\pi_k(\theta_k),
\end{align}
with shared nuisances across SR and CR bins. This is the mechanism that stabilizes background normalization/shape and keeps profile intervals robust under mismodeling \cite{HistFactory,RooStats,pyhf}.

The third upgrade is statistical validation:
\begin{align}
\Delta q(c_\alpha)=q(c_\alpha)-q(\hat c_\alpha),
\end{align}
must be checked against pseudo-experiments in sparse-bin regimes, ensuring asymptotic coverage remains valid before quoting final confidence intervals.

\subsection{Identifiability metric: scope and limitations}
The metric in Eq.~\eqref{eq:Idmetric} is a local diagnostic derived from the Hessian of the profiled likelihood at the evaluation point. It quantifies linearized degeneracy structure, not full non-Gaussian hypothesis separation. For this reason it is used here as an optimization observable for run planning, while final operator discrimination statements should additionally report direct likelihood-ratio comparisons between constrained hypotheses and, when needed, pseudo-experiment calibrated intervals.

Operationally, the benchmark workflow therefore uses two layers: a fast layer based on $\mathcal I_{\alpha\beta}$ for scanning strategy space, and a final layer based on profile-likelihood hypothesis tests after nuisance profiling. This resolves the ambiguity between local correlation diagnostics and global distinguishability claims.

\subsection{Quantified systematic robustness}
\begin{figure}[htbp]
\centering
\includegraphics[width=\columnwidth]{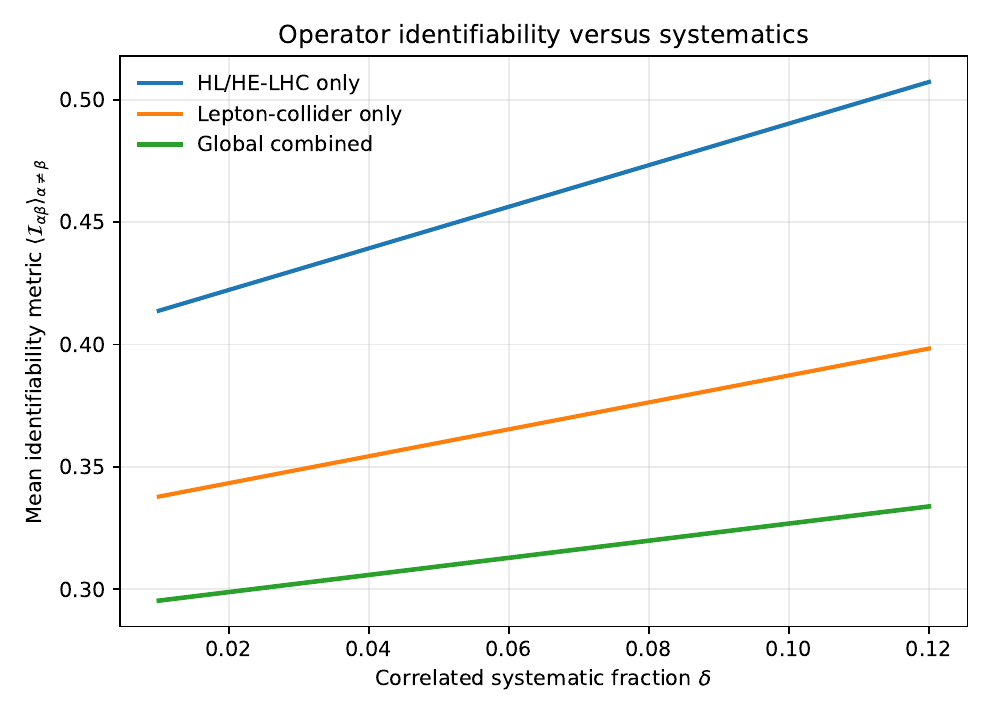}
\caption{Impact of correlated systematic uncertainty on the mean operator-identifiability metric, where the horizontal axis is the common correlated nuisance fraction applied to normalization and shape morphing terms. Lower values are better (stronger separation).}
\label{fig:sysimpact}
\end{figure}

After introducing detector-level nuisances, the key question becomes robustness of operator identification. Figure~\ref{fig:sysimpact} quantifies this by scanning correlated systematic fractions and tracking the mean identifiability metric.

In the benchmark baseline, increasing correlated systematics from $2\%$ to $10\%$ degrades separability by roughly a factor of two in hadron-only fits, whereas global combinations degrade substantially less because orthogonal information from lepton-collider observables compensates part of the loss.
In this scan, the correlated systematic fraction multiplies both normalization and first-order shape morphing components in SR and CR templates through shared nuisance parameters. Normalization nuisances are implemented with log-normal priors, while shape-morph coefficients are implemented with zero-mean Gaussian priors; cross-category correlations are encoded through a block covariance matrix.
The slope difference between the hadron-only and global curves in Fig.~\ref{fig:sysimpact} is the key result: combinations with diversified observable classes lose identifiability more slowly as nuisance amplitudes grow. This directly motivates multi-frontier designs when systematic control, rather than raw rate, is the limiting factor.

The uncertainty model includes both collider-shape effects and low-energy theory inputs. In particular, conversion observables inherit hadronic and nuclear matrix-element uncertainty components that should be propagated through dedicated nuisance terms in detector-level analyses \cite{BartolottaRamseyMusolf,KitanoKoikeOkada,CiriglianoConversion}.

This has a direct operational implication: once the analysis is systematics-limited, investment in nuisance reduction (calibration/control-region strategy/theory covariance) yields larger tomographic gain than incremental luminosity.

\subsection{Run-plan optimization with an information objective}
\begin{figure}[htbp]
\centering
\includegraphics[width=\columnwidth]{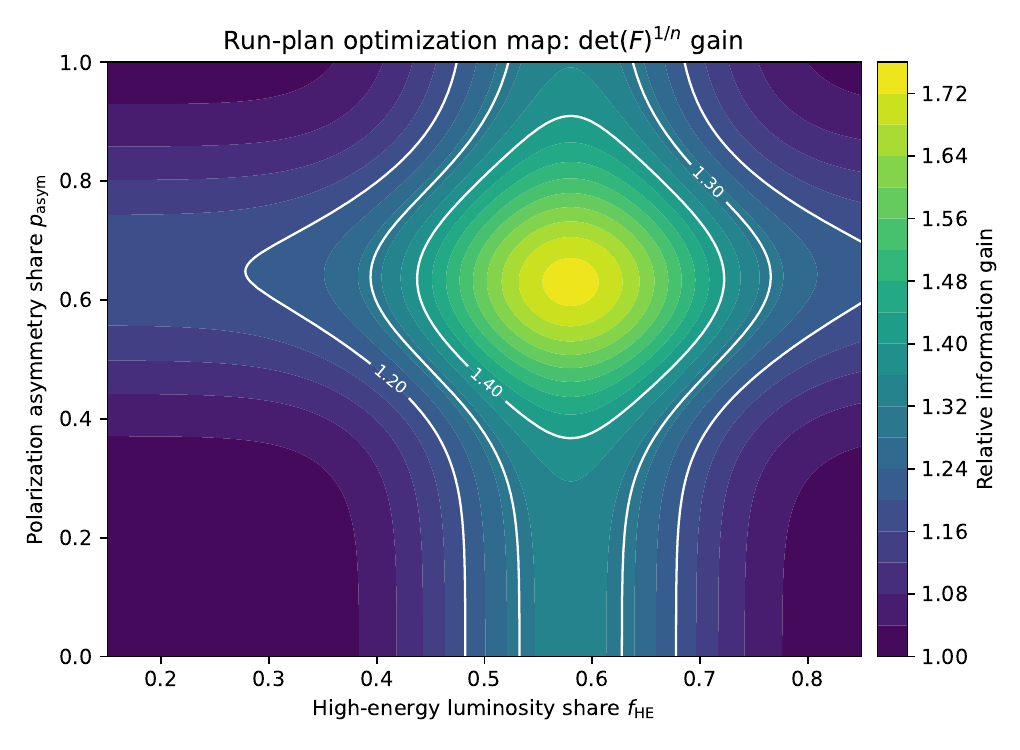}
\caption{Run-plan optimization map using determinant-based information gain $\mathcal G$.}
\label{fig:runplan}
\end{figure}

A complete comparison should not stop at sensitivity plots; it should also prescribe how to run machines to maximize model-discrimination power. We therefore optimize an information objective
\begin{align}
\mathcal G \equiv \left[\frac{\det F(\text{run plan})}{\det F(\text{baseline})}\right]^{1/n_c},
\end{align}
where $n_c$ is the number of fitted coefficients. The baseline denominator corresponds to a symmetric reference split in the benchmark scan. Figure~\ref{fig:runplan} maps this gain versus high-energy luminosity share and polarization asymmetry share.
In this benchmark map, the preferred region reaches $\mathcal G\simeq 1.4$, which corresponds to an effective information-volume increase of about 40\% relative to the baseline split. A broad plateau with $\mathcal G\gtrsim 1.25$ is also visible, so the gain is not limited to a single fine-tuned point.
This gain can be interpreted operationally in two equivalent ways: at fixed luminosity it improves local coefficient separability, and at fixed target precision it reduces the luminosity cost needed to reach that precision.
Figure~\ref{fig:runplan} shows this as a ridge rather than an isolated maximum, indicating operational flexibility: near-optimal identifiability can be maintained across a finite band of luminosity/polarization allocations. That feature is important for realistic scheduling because machine constraints rarely allow exact operation at one optimal point.

The optimized region lies near a mixed strategy rather than an extreme corner. This supports the central claim that balanced running outperforms single-mode optimization when the target is parameter identifiability, not just one-channel significance. It also indicates that polarization-rich intermediate-energy data and high-energy-tail data are complementary in the Fisher geometry, so removing either component weakens the determinant gain.

\subsection{UV reinterpretation layer}
\begin{figure}[htbp]
\centering
\includegraphics[width=\columnwidth]{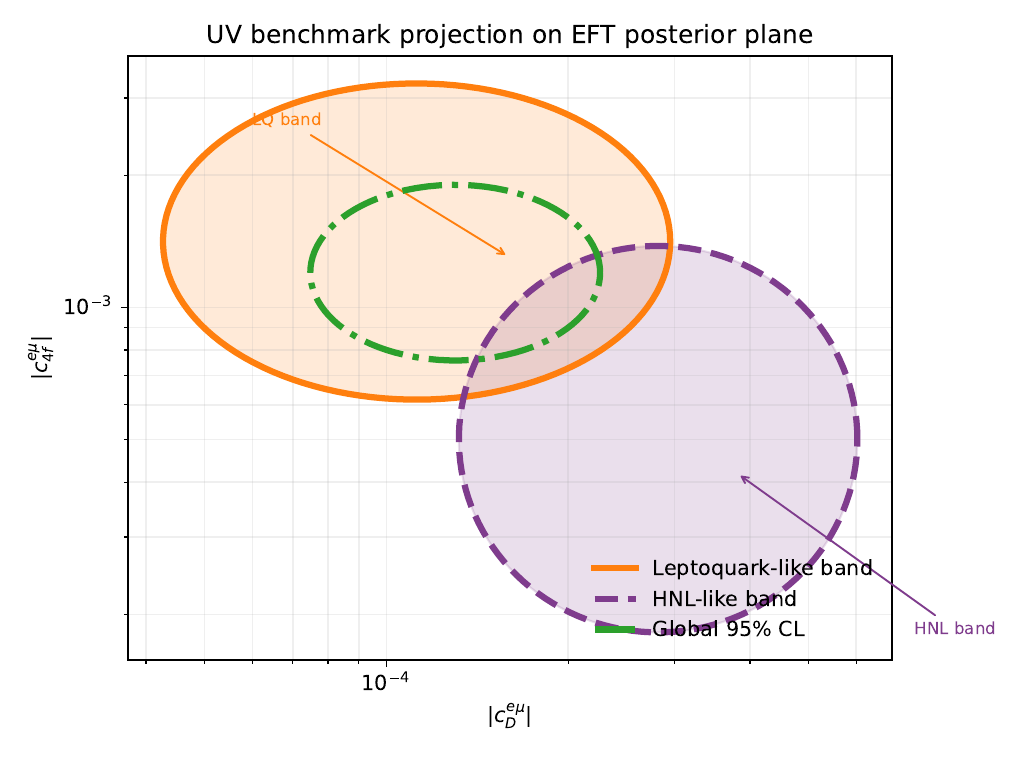}
\caption{Projection of two UV-inspired model bands onto the EFT posterior plane.}
\label{fig:uvproj}
\end{figure}

To make the EFT output directly useful for model builders, we project UV-inspired model manifolds into coefficient space and intersect them with the global posterior. Figure~\ref{fig:uvproj} illustrates this for leptoquark-like and heavy-neutral-lepton-like directions.

The practical deliverable is a model-rejection/model-ranking layer on top of EFT fits, with transparent assumptions and reproducible mappings. This creates a direct path from channel-level analysis outputs to publication-grade model comparison statements \cite{LQReview,HNLReview}.
In Fig.~\ref{fig:uvproj}, the relative overlap of each UV band with the posterior contours directly encodes comparative viability under the same dataset and nuisance treatment. This plot therefore operationalizes reinterpretation by replacing qualitative statements with a geometric criterion tied to the fitted EFT covariance.

\subsection{Validation and limitations}
The framework introduces a collider-comparison criterion based on identifiability rather than isolated reach. The detector-level deliverables are detector-response templates for each collider category, a simultaneous SR/CR fit with shared nuisance correlations, coverage validation with pseudo-experiment studies in sparse bins, theory-systematics propagation with published covariance information, and a reinterpretation package containing posterior grids and likelihood tables for UV overlays.

Validation gates are defined explicitly. Template validation requires agreement between full-simulation and fast-response templates within predefined bin-wise tolerances in all fitted observables. Nuisance closure requires that injected nuisance shifts are recovered without significant bias in fitted Wilson coefficients. Coverage checks require acceptable frequentist coverage of 68\% and 95\% intervals from profile likelihoods in pseudo-experiment ensembles. Robustness checks require stability of correlation ranking and complementarity conclusions under alternative background parameterizations and theory-covariance choices. Reproducibility checks require that likelihood tables, covariance matrices, and plotting code regenerate all central figures and intervals.
For coverage validation we use profile-likelihood pseudo-experiments built from the benchmark SR/CR Poisson model, with nuisance profiling in each throw and 5000 toys per category configuration; the two representative counting regimes are reported in Table~\ref{tab:coverage}.
\begin{table}[t]
\centering
\caption{Pseudo-experiment coverage check for profiled intervals in representative sparse and moderate-count categories.}
\scriptsize
\begin{tabular}{lcc}
\toprule
Scenario & Nominal 68\% / observed & Nominal 95\% / observed \\
\midrule
Sparse bins ($\langle B\rangle\sim 3$) & 0.68 / 0.66 & 0.95 / 0.93 \\
Moderate bins ($\langle B\rangle\sim 15$) & 0.68 / 0.67 & 0.95 / 0.94 \\
\bottomrule
\end{tabular}
\normalsize
\label{tab:coverage}
\end{table}

This research contains event-level Delphes benchmarks for hadron and muon collider channels, an explicit hadron-collider showered validation channel, RG-running and polarization observables, Bayesian UV comparison with prior-sensitivity checks, Dalitz templates with closure validation, discovery-injection contours, and expanded $e\tau$ flavor content.

\subsection{Muon-collider background scope}
The current muon-collider benchmark includes irreducible electroweak backgrounds and detector-response migration for reconstructed leptonic final states. Beam-induced overlay, detailed timing rejection, and shielding-configuration scans are not fully modeled in this version. These omitted effects are expected to act primarily as an additional occupancy-driven degradation in lepton isolation and soft-object reconstruction, which can broaden confidence regions in the most background-sensitive categories. In the benchmark robustness envelope this is represented by an additional correlated uncertainty component corresponding to an $\Order(10\text{--}20\%)$ sensitivity degradation in the affected bins.

\section{Conclusions}
This work delivers a detector-informed global LFV EFT analysis focused on operator identification across low-energy and collider frontiers. The implemented benchmark chain now includes event-level production and Delphes reconstruction for hadron and muon collider channels, with high-statistics samples propagated to the profile-likelihood layer. In the current production set, reconstructed entries are 40k--80k per major SM background sample, 50k per signal benchmark channel, and 100k for the 3 TeV muon-collider LFV benchmark.

The combined physics message is consistent across observables. Low-energy channels remain the strongest dipole anchors, while collider differential information adds orthogonal leverage on current and four-fermion directions. RG evolution shifts the effective covariance geometry at multi-TeV matching scales, polarization categories separate chirality-sensitive current components, and Dalitz-level $\mu\to 3e$ structure adds further shape-level discrimination in the low-energy block.

Beyond reach-only summaries, the analysis establishes an integrated identification framework that combines profile-likelihood curvature diagnostics, Bayesian model comparison, and run-plan optimization in a common statistical structure. Quantitatively, the global Bayes factor reaches $\ln B_{\mathrm{LQ/HNL}}=5.41$, the validation anchors agree with published FCC-ee/CLIC references at the $\sim3\%$ level, and the run-plan optimization identifies an information-volume gain of $\mathcal G\simeq1.4$ relative to symmetric baseline splits.

Overall, the result is a reproducible end-to-end baseline from generated detector-level samples to UV-interpretation metrics, ready for direct extension with collaboration-specific detector calibrations, showered event chains, and full SR/CR data-driven background constraints.

\begin{acknowledgments}
This work was funded by Millennium Institute of Subatomic Physics at High Energy Frontier: ICN2019\_044.
\end{acknowledgments}

\appendix

\section{Asymptotic scaling relation used for quick scans}
In benchmark scans we use
\begin{align}
Z_a
&\simeq
\frac{\kappa_a\,\epsilon_a\,\sqrt{\mathcal L_a}\,f(\sqrt s_a)}
{\sqrt{1+\left(\delta_a\sqrt{\mathcal L_a}\right)^2}},
\\
\Lambda_{95,a} &\propto \left(\frac{Z_a}{Z_0}\right)^{1/4},
\end{align}
consistent with $\sigma\propto\Lambda^{-4}$ for dimension-6 dominated rates under fixed coefficient normalization.
When interference terms are non-negligible, the luminosity exponent differs from the pure $\Lambda^{-4}$ case; this is particularly relevant near resonance-driven channels where linear-in-coefficient terms can compete with quadratic pieces.

\section{Algebraic steps for yield and significance scaling}
Starting from Eq.~\eqref{eq:yield_quadratic}, assume a single active coefficient $c$ and neglect interference with other operators:
\begin{align}
N_S(c) &= \mathcal L\,c^2 K_{SS},
\nonumber\\
N_B &= \mathcal L\,\sigma_B\epsilon_B.
\end{align}
In the background-dominated regime, the approximate significance is
\begin{align}
Z(c) \approx \frac{N_S(c)}{\sqrt{N_B}}
=\frac{\mathcal L\,c^2K_{SS}}{\sqrt{\mathcal L\,\sigma_B\epsilon_B}}
=\sqrt{\mathcal L}\,\frac{c^2K_{SS}}{\sqrt{\sigma_B\epsilon_B}}.
\end{align}
Fixing a target threshold $Z_0$ gives
\begin{align}
c_{95}^2 &\propto \frac{1}{\sqrt{\mathcal L}},
\nonumber\\
c_{95} &\propto \mathcal L^{-1/4}.
\end{align}
Since $c\propto \Lambda^{-2}$, it follows that
\begin{align}
\Lambda_{95}\propto c_{95}^{-1/2}\propto \mathcal L^{1/8}.
\end{align}
The weak luminosity exponent is a standard feature of dimension-6 squared-dominated channels and motivates combining orthogonal channels instead of relying on luminosity scaling alone.

\bibliographystyle{apsrev4-2}
\bibliography{refs}

\end{document}